\documentclass{ws-ijmpa}
\newcommand{\be}{\begin{equation}}
\newcommand{\ee}{\end{equation}}
\def\bea{\begin{eqnarray}}  
\def\eea{\end{eqnarray}}

\def\ie{{\it i.e.}}
\def \GeV{{\mathrm{GeV}}}

\def\LEP2{{LEPII}}
\def\Frac#1#2{\frac{\displaystyle{#1}}{\displaystyle{#2}}}
\def\lsim{\raise0.3ex\hbox{$\;<$\kern-0.75em\raise-1.1ex\hbox{$\sim\;$}}}
\def\gsim{\raise0.3ex\hbox{$\;>$\kern-0.75em\raise-1.1ex\hbox{$\sim\;$}}}
\def\no{\nonumber\\}

\def\neq{\not=}
\def \av#1{\left\langle #1\right\rangle}

\def \GeV{{\mathrm{GeV}}}

\def \Tr{{\mathrm{Tr}}\,}
\def \av#1{\left\langle #1\right\rangle}

\def\plb#1#2#3{    {\it Phys. Lett. }{\bf B #1} (#2) #3}
\def\prd#1#2#3{    {\it Phys. Rev. }{\bf D #1} (#2) #3}

\def\prl#1#2#3{    {\it Phys. Rev. Lett. }{\bf #1} (#2) #3}

\begin{document}

\title{CP violation in supersymmetric theories}

\author{ Shaaban Khalil}

\address{IPPP, Physics Department, Durham University, DH1 3LE,
Durham, U.K\\
Ain Shams University, Faculty of Science, Cairo, 11566, Egypt.}

\maketitle

\begin{abstract}
We review the present status of CP violating problem in supersymmetric extensions
of the standard model. We analyze the constraints imposed by the experimental 
limits of the electron, neutron, and mercury electric dipole moments on the 
supersymmetric CP phases and show that only the scenarios with flavour-off-
diagonal CP violation remain attractive. These scenarios require hermitian 
Yukawa matrices which naturally arise in models with left--right symmetry or 
a $SU(3)$ flavour symmetry. In this case, $\varepsilon_K$  and 
$\varepsilon'/\varepsilon$ can be saturated by a small non--universality of 
the soft scalar masses through the gluino and chargino contributions 
respectively. The model also predicts a strong correlation between 
$A_{CP}(b\to s \gamma)$ and the neutron electric dipole moment. 
In this framework, the standard model gives a the leading contribution to 
the CP asymmetry in $B \to \psi K_S$ decay, while the dominant chargino 
contribution to this asymmetry is $< 0.2$. Thus, no constraint is set on 
the non--universality of this model by the recent BaBar and Belle 
measurements. 
\end{abstract}

\section{Introduction}
The understanding of the origin of CP violation is one of the remanining 
open questions in particle physics.  In the standard model (SM), CP violation 
and flavour transition arise from the complex Yukawa couplings which have a 
Cabibbo--Kobayashi--Maskawa (CKM) mixing matrix with physical $\delta _{CKM}$ 
phase of order unity. 
At present, CP violation is observed in the $K$ and $B$ systems and the
experimental results are consistent with the SM.
However, CP violation can be tested from the existing bounds on the electric 
dipole moment (EDM) of the neutron and the electron. It is remarkable that the 
SM contribution to the EDM of the neutron is of order $10^{-30}$ e.cm, while 
the experimental limit is of order $10^{-26}$ e cm \cite{edmexp}. 
We expect that with the 
further improvements of experimental precision the EDM will provide a crucial 
test of CP violation. Also more experimental information on the CP violation 
will be obtained from the B-factory soon, which would be a crucial test of the 
CP violation in SM. Furthermore, SM model is unable to explain the 
cosmological baryon asymmetry of our universe, the presence of new sources 
of CP violation is required for this explanation.

In supersymmetric (SUSY) extensions of the SM there are additional sources of
CP violation, due to the presence of new CP violating phases
which arise from the complexity of the soft SUSY breaking terms and
the SUSY preserving $\mu$-parameter. These new phases have significant
implications and can modify the SM predictions in CP violating phenomena.
In particular, they would give large contributions to the 
electric dipole moment (EDM) of the electron, neutron and mercury atom~\cite{AKL},
to CP violating parameters $(\varepsilon_K, \varepsilon'/\varepsilon)$ of 
$K-\bar{K}$ system \cite{susyK}, and to the CP asymmetries in the $B-\bar{B}$ 
system \cite{susyB}. 
These phases can be classified into two categories.
The first category includes  flavor-independent phases such as the phases
of the $\mu$-parameter, $B$-parameter, gaugino masses and the overall
phase of the trilinear couplings. The other category includes
the flavor-dependent phases, i.e. the phases of the off--diagonal elements
of $A_{ij}$ after the overall phase is factored out and phases in the squark 
mass matrix $ m_{ij}^2$. Two of the flavor-independent phases can be eliminated 
by the $U(1)_R$ and $U(1)_{PQ}$ transformations under which these parameters
behave as spurions. The Peccei--Quinn transformation act 
on the Higgs doublets and the right--handed superfields in such a way that 
all the interactions but which mix the two doublets are invariant. The 
Peccei--Quinn charges are $Q_{PQ}(\mu)=Q_{PQ}(B\mu)$, 
$Q_{A}(\mu)=Q_{PQ}(m_i)=0$. The $U(1)_R$ trnasforms the Grassmann variable
$\theta \to \theta e^{i \alpha}$ and the fields in such a way that the 
integral of the superpotential over the Grassmann variables is invariant,
{\it i.e.,} the $U(1)_R$ charge of the superpotential is 2. As a result,
$Q_R(B\mu) = Q_R(\mu) -2$, $Q_R(A) = Q_R(m_i) =-2$. The six physical CP--phases
of the theory are invariant under both $U(1)_R$ and $U(1)_{PQ}$, and 
can be chosen as 
\be
\mathrm{Arg}\left(A^*_d m_i\right),  \hspace{2cm} \mathrm{Arg}\left((B\mu)^* 
\mu A_{\alpha}\right),
\ee
where $i=1,2,3$ and $\alpha=d,u,l$. All other CP--phases can be expressed as
linear combinations. If the $A$--terms and the gaugino masses are universal, 
there are two physical phases $\mathrm{Arg}(A^* m)$, 
$\mathrm{Arg}(B^* A)$. 

However, the non--observation of EDMs imposes a stringent constraint on
flavor--independent SUSY phases. Putting these new phases to zero is not 
natural in the sense of 't Hooft~\cite{hooft} since the Lagrangian does not 
acquire any new symmetry in the limit where these new phases vanish. 
This is the so-called SUSY CP-problem. 
In this article we will review the constraints imposed by the EDMs on the
flavor-diagonal CP-phases and the possible scenarios allowing to supress the EDM contributions: 
small SUSY CP phases, heavy sfermions, EDM cancellation, and flavor
off--diagonal CP violation. Also we will discuss  aspects of  CP
violation in the $K$ and $B$ systems due to the flavor-off-diagonal phases from
non-degenerate $A$--terms.

This article is organized in the following way. In section 2 we 
briefly discuss the origin of the soft SUSY breaking terms.
Section 3 is devoted to the study of the constraints from the EDMs of the 
electron, neutron, and mercury atom in generic SUSY models.
We present the possible ways to avoid overproduction of EDMs and explain 
that, in SUSY models with flavor off--diagonal phases, the EDMs can be 
kept sufficiently small while these phases unconstrained. The effect 
of the off--diagonal phases in the CP violation process in the kaon system
is given in section 4. In section 5 we analyze the CP violation in the 
$B$--sector and show that in this framework the large CP asymmetry in the 
$B\to \psi K_S$ decay is given by the SM contribution while the 
SUSY contribution is very small. In contrast, the SUSY contribution to the 
CP asymmetry in the $B\to X_S \gamma$ decay can be as large as $\pm 10\%$.
Finally, the conclusions are presented in section 6.

\section{The origin of the soft terms}
It is clear that the SUSY CP violating problem is a problem of SUSY breaking 
since the relevant phases originate from SUSY breaking terms.
In this section we briefly discuss the possible
mechanisms which may give rise to the SUSY soft breaking terms. The general
structure one has in mind includes three sectors:\\
$i)$ The observable sector which comprises all the ordinary particles and
their SUSY partners,\\
$ii)$ a ``hidden" or ``secluded" sector where the breaking of SUSY occurs,\\
$iii)$ the messengers of the SUSY breaking from the hidden to observable
sector. 

        The two most explored alternatives that have been studied in this
context are:\\
$a)$ SUSY breaking supergravity where the mediators are gravitational
interactions, the scale of SUSY breaking is $M_S \simeq  \sqrt{M_P\ M_W}$
and the mass of the fermionic partner of the gravition, the gravitino, is
$m_{3/2} \propto M_S^2/M_P$.\\
$b)$ SUSY broken at a much lower scale with messengers provided by some
gauge interactions. In this case $m_{3/2} = M_S^2/M_P$ is generally
very small, the gravitino being the most likely candidate for the LSP of
the theory.

        We start the former and still more popular alternative. As well known 
that the invariance under local SUSY transformation implies invariance under 
local coordinate change. Thus, local SUSY (supergravity) naturally includes 
gravity. The effective action, up to two derivatives, is completely specified
in terms of three functions which depend on the chiral superfields of the
theory. The K\"ahler potential $K(\Phi, \Phi^{\dag})$, the analytic
superpotential $W(\Phi)$ and the gauge kinetic functions $f^a(\Phi)$ which
are also analytic functions of the chiral superfields. The tree level
supergravity scalar potential is given by~\cite{sgravity}
\begin{equation}
V= e^{G} \left[ G^i (G^{-1})^j_i G_j -3\right] +\frac{1}{2}
f^{-1}_{\alpha \beta} D^{\alpha} D^{\beta},
\label{vsupergravity}
\end{equation}
where $G(\Phi, \Phi^*)= K(\Phi, \Phi^*) + \log \vert W(\Phi)\vert^2$,
$G^i = \frac{\partial G}{\partial \phi_i^*}$ and $G_i^j= \frac{\partial G}
{\partial \phi^i \partial \phi_j^*}$. The gravitino mass is given by
Eq.\ref{vsupergravity} shows that unlike the case of the global supersymmetry
one can simultaneously have $\langle G_i \rangle \neq 0$ \footnote{
The auxiliary fields of the chiral and vector supermultiplets are given in
terms of $G_i$ see Ref.\cite{sgravity}. One can see that if at least one of
the $G_i$ VEV's is non-vanishing SUSY is broken} (which breaks SUSY) and 
vanishing cosmological constant.

        Supergravity is broken in a ``hidden sector", namely a sector of the
theory that couples to the ``observable sector" of quarks, leptons, gauge
fields, Higgs and their supersymmetric partners, only through gravitational
interactions. The simplest model of a supersymmetry breaking hidden sector
is known as ``Polonyi model". In this model the minimal form of the K\"ahler
potential $K= z^{*i} z_i + y^{*r} y_r $ is assumed ($z_i$ and $y^r$ denote
the fields of hidden and observable sectors respectively). The resulting
soft breaking terms are obtained taking the so-called flat limit , \ie\
sending $M_P \rightarrow \infty$ while keeping the ratio
$M_S^2/M_P= m_{3/2}$ fixed ( $M_S$ denotes the scale of supergravity
breaking). One obtains a common mass term for all the scalar particles in
the observable sector which is equal to the gravitino mass. Also in this
model we get a universal $A$ term (\ie\ $A_U=A_D=A_L=A$). 

        However, it is possible to obtain effective potentials in which 
this universality is absent by taking the kinetic terms for the chiral 
superfields to non-minimal. As recently stresse, the soft supesymmetry 
breaking parameters may be non-universal in the effective theories which 
are derived from the superstring theories. In general supergravity, the soft
scalar mass and the $A$-parameter are given by
\begin{eqnarray}
m^2_{\alpha} &=& m^2_{3/2} - \bar{F}^{\bar{m}} F^n \partial_{\bar{m}} \partial_n
\ln \tilde{K}_{\alpha},
\label{scalar}\\
A_{\alpha \beta \gamma} &=& F^m \Big[ \hat{K}_m + \partial_m \ln Y_{\alpha \beta
\gamma} -\partial_m \ln(\tilde{K}_{\alpha} \tilde{K}_{\beta}
\tilde{K}_{\gamma})\Big],
\label{trilinear}
\end{eqnarray}
where $m$ refers to the SUSY breaking fields and 
$\tilde{K}_{\alpha} \equiv \tilde{K}_{\bar{\alpha} \alpha}$.
Note that in Eq.(\ref{scalar}) we 
have diagonal soft scalar masses due to the assumption that the K\"ahler 
metric is diagonal. General supergravity models allow for a non-vanishing 
off-diagonal K\"{a}hler metric $K_{\bar{\alpha} \beta}$. However, such mixing typically
does not appear in superstring models at the leading order due to 
additional ``stringy'' symmetries beyond those of the Standard Model.
The gaugino masses are given in terms of the gauge kinetic
function $f_a$, where $\mathrm{Re} f_a = 1/g_a^2$ and the
subscript $a$ represents the corresponding gauge group,
\be 
M_a = \frac{1}{2} (\mathrm{Re} f_a)^{-1} F^m
\partial_m f_a \;. 
\ee 
Apart from $m_{\alpha}$, $A_{\alpha \beta \gamma}$, and $M_a$ soft parameters,
soft bosonic bilinear can be also present. In case of the MSSM, all the symmetries
of the low energy allow for a superpotential term coupling the two Higgs
doublets of the form $W=\mu H_1 H_2$. The associated soft breaking term in the
scalar potential will have the form $B\mu H_1 H_2 + h.c.$, where $B$ is a 
dimensionful parameter of order gravitino mass. It is well known that in order
to get appropriate $SU(2)_L \times U(1)Y$ breaking, the $\mu$ has to be of the
same order of magnitude as the SUSY breaking soft terms.

	Now, we turn to the alternative mechanism of breaking 
supersymmetry at low energies and that gauge interactions are the
``messenger" of supersymmetry breaking~\cite{gmsb}. The gauge mediated 
supersymmetry breaking (GMSB) models have several attractive features.
As gauge interactions are flavor blind, squark and slepton masses are
universal. In the minimal model of this kind, the messanger fields 
transform as a single flavor of $5+\bar{5}$ of $SU(5)$. Hence, there 
are $SU(2)_L$ doublets $l$ and $\bar{l}$ and $SU(3)_C$ triplets $q$ and
$\bar{q}$. In order to introduce supersymetry breaking into the 
messanger sector, these fields may be coupled to a gauge singlet spurion,
$S$, through the superpotential
\begin{equation}
W= \lambda_1 S l \bar{l} + \lambda_2 S q \bar{q},
\end{equation}
due to interactions with some supersymmetry breaking sector of theory. The
field $S$ has a non-zero expectation value both for its scalar and auxiliary
component, $\langle S \rangle$ and $\langle F_S \rangle$. Integrating
out the messenger sector give rise to gaugino masses at one loop.
\begin{equation}
m_{\lambda_i}= c_i \frac{\alpha_i}{4\pi} \Lambda,
\end{equation}
where $\Lambda=\frac{\langle F_S \rangle}{\langle S \rangle}$, $c_1=5/3$,
$c_2=c_3=1$ and $\alpha_1=\frac{\alpha}{\cos^2\theta_W}$. For the scalar
masses one has
\begin{equation}
\tilde{m}^2 = 2 \Lambda^2 \left[ C_3 (\frac{\alpha_3}{4\pi})^2 +
C_2 (\frac{\alpha_2}{4\pi})^2+\frac{5}{3} (\frac{Y}{2})^2
(\frac{\alpha_1}{4\pi})^2 \right],
\end{equation}
where $C_3=4/3$ for color triplets and zero for singlets, $C_2= 3/4$ for
weak doublets and zero for singlets, and $Y=2(Q-T_3)$ is the ordinary
hypercharge. In this model, the $A$-terms are high order effects. The $\mu$
and $B$ parameters can arise through interactions of the Higgs field with
various singlets~\cite{dvali}. This model has few new parameters relative
to the minimal standard model. These parameters are $\Lambda$, $\mu$ and
$B$. Since the scalar masses are functions of only gauge quantum numbers,
these models also automatically solve the supersymmetric flavor problem.

The most important difference between the models of the gravity mediated
SUSY breaking and the models of gauge mediation is that in the later models
the LSP is the gravitino and its mass is of order $10^{-10}$ GeV. Thus
leads to a very different phenomenology from the formare model. It allows
the lightest neutralino $\chi_1^0$ to decay to photon plus gravitino and
this gives a signal for SUSY as a missing energy.

\section{SUSY CP violating phases and constraints from the EDMs}
As mentioned in the introduction, in supersymmetric theories there are several 
possible sources for CP violating phases in addition to the CKM phase. 
These CP violating phases have an important impact on the
phenomenology of CP violatin. They can induce an EDM of quarks and leptons
at one-loop level. The EDMs of the neutron, electron, 
and mercury atom \cite{edmexp} impose severe constraints on the flavor-diagonal 
phases, the so--called SUSY CP problem.
\begin{eqnarray} 
d_n &<& 6.3 \times 10^{-26} ~\mathrm{e~cm}~, \label{nedm:bound}\\
d_e &<& 4.3 \times 10^{-27} ~\mathrm{e~cm}~,\\
d_{Hg} &<& 2.1 \times 10^{-28} ~\mathrm{e~cm}~.
\label{mercury:bound}
\end{eqnarray}
With the expected improvements in experimental precision,  
the EDM is likely to be one of the most important tests for
physics beyond the Standard Model for some time to come, and EDMs will 
remain a difficult hurdle for supersymmetric theories. 
Indeed it is remarkable that the SM contribution to the EDM of the
neutron is of order $10^{-30}$ e cm, whereas the ``generic'' 
supersymmetric value is $10^{-22}$e cm. 

In this section we review EDM constraints in the context of  supersymmetric 
theories as well as  known mechanisms to suppress the EDMs.
We also study to what extent different scenarios rely on assumptions 
about the neutron structure, i.e. chiral quark model {\em vs} parton model. 
Let us first summarize the contributions to the three most  
significant EDMs, beginning with the most reliable, the electron EDM. 

The electron EDM is defined by the effective CP-violating interaction
\begin{equation}
{\cal{L}}= -{i\over 2} d_e \bar e \sigma_{\mu\nu}\gamma_5 e \;F^{\mu\nu}\;,
\end{equation}
where $F^{\mu\nu}$ is the electromagnetic field strength.
The experimental bound on the electron EDM is derived from the electric dipole
moment of the thallium atom and is given by
\begin{equation}
d_e < 4 \times 10^{-27} {\rm e\;  cm}\;.
\end{equation}
In supersymmetric models, the electron EDM arises due to CP-violating 1-loop 
diagrams with the chargino and neutralino exchange:
\begin{equation}
d_e=d_e^{\chi^+}+d_e^{\chi^0}\;.
\end{equation}
Since the EEDM calculation involves little uncertainty it allows to extract 
reliable bounds on the CP-violating SUSY phases.

The neutron EDM has contributions from a number of CP-violating operators 
involving quarks, gluons, and photons. The most important ones include the 
electric and chromoelectric dipole operators, and the Weinberg three-gluon 
operator:
\begin{eqnarray}
 {\cal{L}}=&-& {i\over 2} d_q^E \bar q \sigma_{\mu\nu}\gamma_5 q \;F^{\mu\nu}\;
- \; {i\over 2} d_q^C \bar q \sigma_{\mu\nu}\gamma_5 T^a q \;G^{a \mu\nu }\; \nonumber\\
&-& \; {1\over 6} d^G f_{abc} G_{a \mu \rho} G_{b \nu}^{\rho}
G_{c \lambda \sigma} \epsilon^{\mu\nu\lambda\sigma}\;,
\end{eqnarray} 
where $G_{a \mu\nu} $ is the gluon field strength, $T^a$ and $f_{abc}$ are the
SU(3) generators and group structure coefficients, respectively.
Given these operators, it is however a nontrivial task to evaluate the neutron EDM
since  assumptions about the neutron internal structure are necessary. 
In what follows we will study two models, namely the quark chiral model and the 
quark parton model. Neither of these models is sufficiently reliable by itself 
\cite{pospelov}, however a power of the combined analysis should provide an 
insight into implications of the bound on the neutron EDM. 
Another approach to the neutron EDM based on the QCD sum rules  has  appeared 
in \cite{pospelov1} and earlier work \cite{khrip1}, \cite{khrip}. 
We note that in any case the NEDM calculations involve uncertain hadronic 
parameters such as the quark masses and thus these calculations have a status 
of estimates. The major conclusions of the present work are independent of the 
specifics of the neutron model.

The chiral quark model is a nonrelativistic model which relates the
neutron EDM to the EDMs of the valence quarks with the help of the SU(6)
coefficients:
\begin{equation}
d_n={4\over 3} d_d -{1\over 3} d_u \;.
\end{equation} 
The quark EDMs can be $estimated$  via Naive Dimensional Analysis \cite{nda} as
\begin{equation}
d_q=\eta^E d_q^E + \eta^C {e\over 4\pi} d_q^C + \eta^G {e \Lambda \over 4\pi} d^G\;,
\end{equation}
where the QCD correction factors are given by $\eta^E=1.53$, $\eta^C \simeq \eta^G
\simeq 3.4$, and $\Lambda \simeq 1.19\;GeV$ is the chiral symmetry breaking scale.
We use the numerical values for these coefficients as given in \cite{nath}.
The parameters $\eta^{C,G}$ involve  considerable uncertainties steming from
the fact that the strong coupling constant at low energies is unknown.
Another weak side of the model is that it neglects the sea quark contributions 
which play an important role in the nucleon spin structure. 

The supersymmetric contributions to the dipole moments of the individual quarks
result from the 1-loop gluino, chargino, neutralino exchange diagrams
\begin{equation}
d_q^{E,C}=d_q^{\tilde g \;(E,C)}+d_q^{\chi^+ \; (E,C)} + d_q^{\chi^0 \; (E,C)}\;,
\end{equation}
and from the 2-loop gluino-quark-squark diagrams which generate $d^G$. 

The parton quark model is based on the isospin symmetry and
known contributions of different quarks to the spin of the proton \cite{ellis}. 
The quantities $\Delta_q$ defined as $\langle n \vert {1\over 2}\bar q \gamma_{\mu}
\gamma_5 q \vert n \rangle = \Delta_q\; S_{\mu} $, where $S_{\mu}$ is the neutron
spin, are related by the isospin symmetry to the quantities $(\Delta_q)_p$ which
are measured in the deep inelastic scattering (and other) experiments, i.e.
$\Delta_u = (\Delta_d)_p$, $\Delta_d = (\Delta_u)_p$, and $\Delta_s = 
(\Delta_s)_p$. To be exact, the neutron EDM depends on the (yet unknown) tensor 
charges rather than these axial charges. The main  $assumption$ of the model is 
that the quark contributions to the NEDM are weighted by the same factors 
$\Delta_i$, i.e.  \cite{ellis}
\begin{equation}
d_n=\eta^E (\Delta_d d_d^E + \Delta_u d_u^E +\Delta_s d_s^E)\;.
\end{equation}
In our numerical analysis we use the following values for these quantities
$\Delta_d=0.746$, $\Delta_u=-0.508$, and $\Delta_s=-0.226$ as they appear 
in the analysis of Ref.\cite{bartl}. 
As before, we have
\begin{equation}
d_q^{E}=d_q^{\tilde g \;(E)}+d_q^{\chi^+ \; (E)} + d_q^{\chi^0 \; (E)}\;.
\end{equation}
The major difference from the chiral quark model is a large strange quark
contribution (which is likely to be an overestimate \cite{pospelov}).
In particular, due to the large strange and charm quark masses,
the strange quark contribution dominates in most regions of the parameter space. 
This leads to considerable numerical differences between the predictions of the two
models.

The EDM of the mercury atom results mostly from T-odd nuclear forces in the 
mercury nucleus, which induce the effective interaction  of the type 
$({\rm {\bf I}} \cdot \nabla ) \delta ( {\rm {\bf r}})$
between the electron and the nucleus of spin {\bf I} \cite{pospelov}.
In turn, the T-odd nuclear forces arise  due to the effective 4-fermion interaction
$\bar p p \bar n i \gamma_5 n$. It has been argued \cite{pospelov} that the mercury
EDM is primarily sensitive to the chromoelectric dipole moments of the quarks
and the limit 
\begin{equation}
d_{Hg} < 2.1 \times 10^{-28} {\rm e \; cm}
\end{equation}
can be translated into
\begin{equation}
\vert d_d^C - d_u^C -0.012 d_s^C \vert /g_s < 7 \times 10^{-27} {\rm cm}\;,
\label{cedmlimits}
\end{equation}
where $g_s$ is the strong coupling constant.
As in the parton neutron model, there is a considerable strange quark contribution.
The relative coefficients of the  quark contributions in (\ref{cedmlimits}) are 
known better than those for the neutron, however the overall normalization is still
not free of uncertainties \cite{khrip}. 

\begin{figure}[ht]
\epsfig{figure=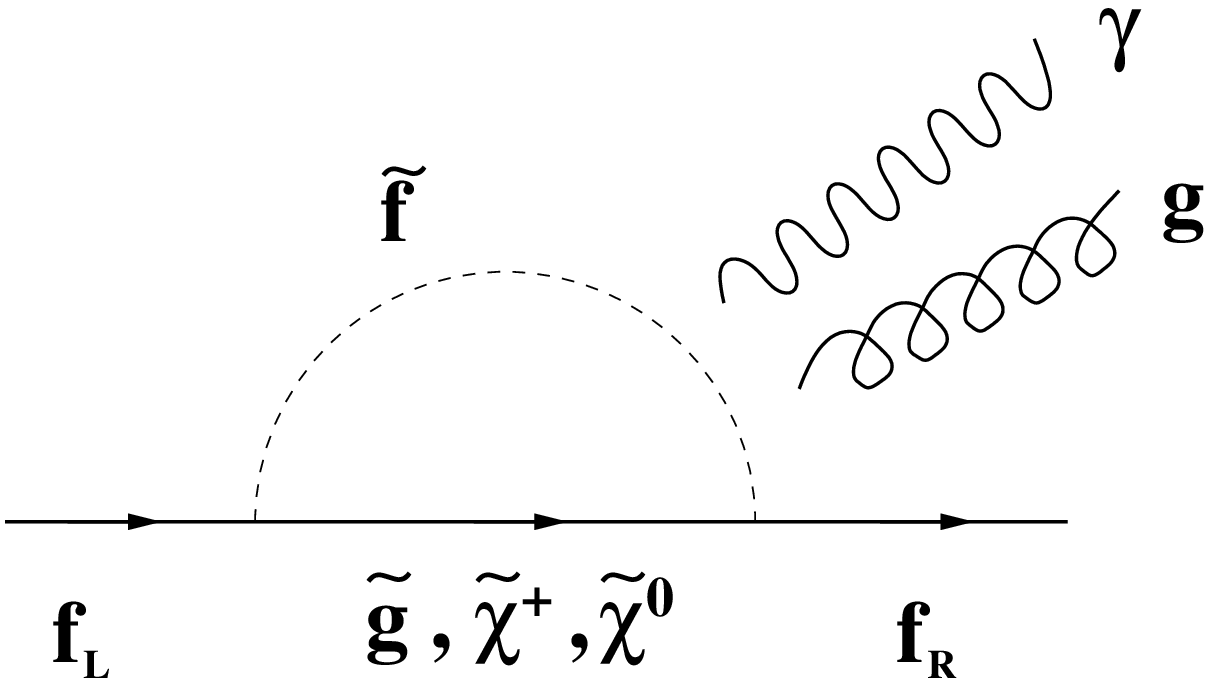,height=4cm,width=5.5cm,angle=0}
\hspace{1.cm}\epsfig{figure=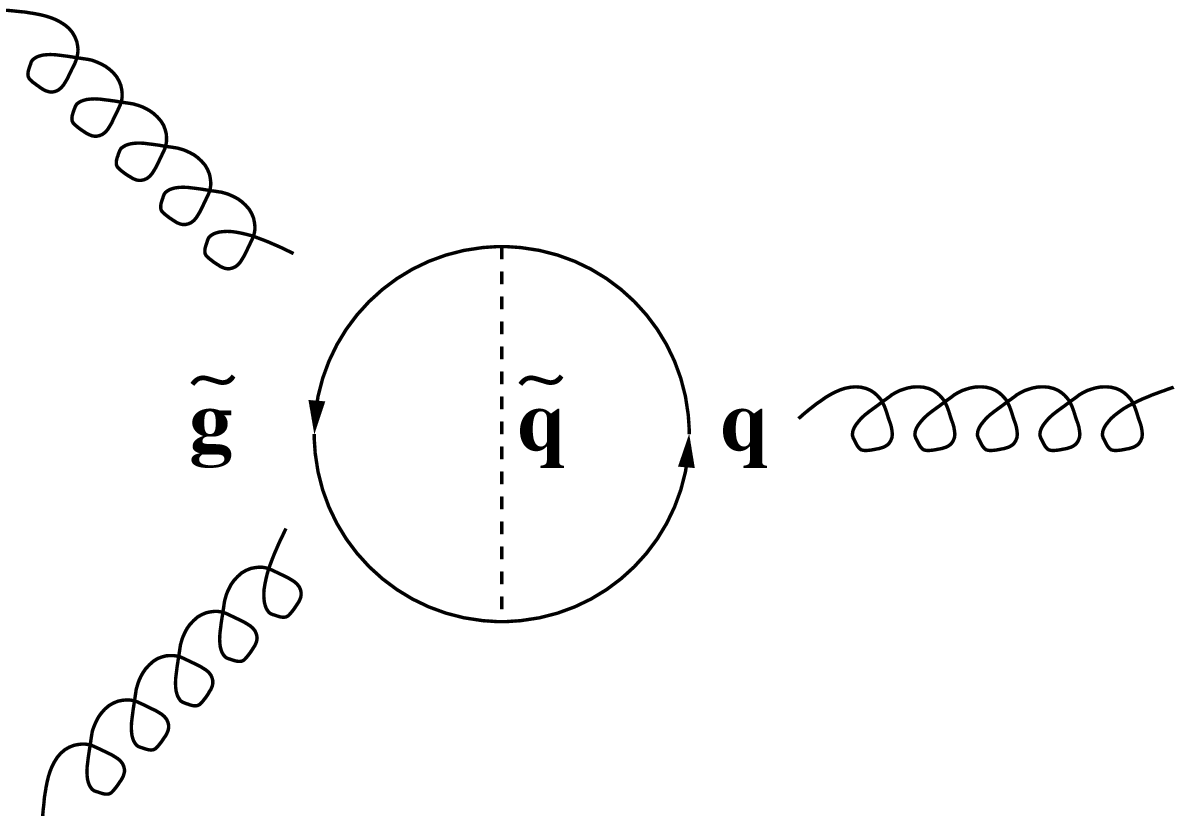,height=4cm,width=5.5cm,angle=0}
\medskip
\caption{Leading SUSY contributions to the EDMs. The photon and gluon lines are to be attached to
the loop in all possible ways.}
\label{diagram1}
\end{figure}

Below we list formulae for individual supersymmetric contributions to the EDMs 
due to the Feynman diagrams in Fig.\ref{diagram1}.
In our presentation we follow the work of Ibrahim and Nath \cite{nath}.

Neglecting the flavor mixing, the electromagnetic contributions to the fermion EDMs are given by \cite{nath}:
\begin{eqnarray}
&& d_q^{\tilde g \; (E)}/e =  \frac{-2 \alpha_{s}}{3 \pi} 
\sum_{k=1}^2  
{\rm Im}(\Gamma_{q}^{1k})  \frac{ M_3}{M_{\tilde{q}_k}^2} Q_{\tilde{q}}\;
 {\rm B}\biggl( \frac{M_3^2}{M_{\tilde{q}_k}^2}\biggr)  \;,\nonumber\\
&& d_u^{\chi^+ \; (E)}/e = \frac{-\alpha_{em}}{4\pi\sin^2\theta_W}\sum_{k=1}^{2}\sum_{i=1}^{2}
      {\rm Im}(\Gamma_{uik})
               \frac{m_{\chi^+_i}}{M_{\tilde{d}_k}^2} \biggl[ Q_{\tilde{d}}\;
                {\rm B} \biggl( \frac{m_{\chi^+_i}^2}{M_{\tilde{d}_k}^2} \biggr)+
        (Q_u-Q_{\tilde{d}})\; {\rm A}\biggl( \frac{m_{\chi^+_i}^2}{M_{\tilde{d}_k}^2}\biggr) \biggr] \;,\nonumber\\
&& d_d^{\chi^+ \; (E)}/e = \frac{-\alpha_{em}}{4\pi\sin^2\theta_W}\sum_{k=1}^{2}\sum_{i=1}^{2}
      {\rm Im}(\Gamma_{dik})
               \frac{m_{\chi^+_i}}{M_{\tilde{u}_k}^2} \biggl[ Q_{\tilde{u}}\;
                {\rm B} \biggl( \frac{m_{\chi^+_i}^2}{M_{\tilde{u}_k}^2} \biggr)+
        (Q_d-Q_{\tilde{u}})\; {\rm A}\biggl( \frac{m_{\chi^+_i}^2}{M_{\tilde{u}_k}^2}\biggr) \biggr] \;,\nonumber\\
&& d_e^{\chi^+}/e=\frac{\alpha_{em}}{4\pi\sin^2\theta_W} 
      \sum_{i=1}^{2} {m_{\chi^+_i} \over  {m_{\tilde{\nu}}^2}} {\rm Im}
     (\Gamma_{ei})\;   
        {\rm A}\biggl( \frac{m_{\chi^+_i}^2}{m_{\tilde{\nu}}^2} \biggr)\;,\nonumber\\ 
&& d_f^{\chi^0 \; (E)}/e =\frac{\alpha_{em}}{4\pi\sin^2\theta_W}\sum_{k=1}^{2}\sum_{i=1}^{4}
{\rm Im}(\eta_{fik})
               \frac{m_{\chi^0_i}}{M_{\tilde{f}_k}^2} Q_{\tilde{f}}\;
{\rm B}\biggl( \frac{m_{\chi^0_i}^2}{M_{\tilde{f}_k}^2}\biggr) \;.
\end{eqnarray}
Here
\begin{equation}
\Gamma_{q}^{1k}=e^{-i\phi_3}D_{q2k}D_{q1k}^* \;,
\end{equation}
with $\phi_3$ being the gluino phase and $D_q$ defined by $D_q^{\dagger} M_{\tilde q}^2 D_q={\rm diag}( M_{\tilde q_1}^2, M_{\tilde q_2}^2)$. The sfermion mass matrix  $M_{\tilde f}^2$ is given by
\be
{\small M_{\tilde{f}}^2\!=\!\left(\!\matrix{{M_L}^2\!+\!m{_f}^2\!+\!M_{z}^2
(\frac{1}{2}\!-\!Q_f\sin^2\theta_W)\cos2\beta\!\!\!&\!\!\! m_f(A_{f}^{*}\!-
\!\mu R_f) \cr m_f(A_{f}\!-\!\mu^{*} R_f)\!\!\!&\!\!\! M_{R}^2\!+\!m{_f}^2\!+
\!M_{z}^2 Q_f \sin^2\theta_W \;\cos2\beta}\!\right)\!,}
\ee
where $R_f=\cot\beta$ $(\tan\beta)$ for $I_3=1/2$ $(-1/2)$.
The chargino vertex $\Gamma_{fik}$ is defined as
\begin{eqnarray}
&&\Gamma_{uik}=\kappa_u V_{i2}^* D_{d1k} (U_{i1}^* D_{d1k}^*-
                \kappa_d U_{i2}^* D_{d2k}^*) \;,\nonumber\\
&&\Gamma_{dik}=\kappa_d U_{i2}^* D_{u1k} (V_{i1}^* D_{u1k}^*-
                \kappa_u V_{i2}^* D_{u2k}^*)
\end{eqnarray}
and analogously for the electron; here $U$ and $V$ are the unitary matrices diagonalizing the chargino mass
matrix: $U^* M_{\chi^+} V^{-1}= {\rm diag} (m_{\chi^+_1},m_{\chi^+_2})$. The quantities $\kappa_f$ are
the Yukawa couplings
\begin{equation}
\kappa_u=\frac{m_u}{\sqrt{2} m_W \sin\beta}, 
 ~~\kappa_{d,e}=\frac{m_{d,e}}{\sqrt{2} m_W \cos\beta}.
\end{equation}
The neutralino vertex $\eta_{fik}$ is given by
\begin{eqnarray}
\eta_{fik} & = &{\biggl[-\sqrt{2} \{\tan\theta_W (Q_f-I_{3_f}) X_{1i}
  +I_{3_f} X_{2i}\}D_{f1k}^*-
     \kappa_{f} X_{bi} D_{f2k}^*\biggr]}\nonumber\\
 &\times& {\biggl[ \sqrt{2} \tan\theta_W Q_f X_{1i} D_{f2k}
     -\kappa_{f} X_{bi} D_{f1k}\biggr]}\;,
\end{eqnarray}
where $I_3$ is the third component of the isospin,
$b=3\;(4)$ for $I_3=-1/2\;(1/2)$, and $X$ is the unitary matrix diagonalizing the 
neutralino mass matrix:  $X^T M_{\chi^0} X= {\rm diag} (m_{\chi^0_1},m_{\chi^0_2},m_{\chi^0_3},m_{\chi^0_4})$. In our convention the mass matrix eigenvalues are  positive and ordered as
$m_{\chi^0_1} > m_{\chi^0_2}>...$ (this holds for all mass matrices in the paper).
The loop functions $A(r), B(r)$, and $C(r)$ are defined by
\begin{eqnarray}
&& A(r)=\frac{1}{2(1-r)^2}\biggl(3-r+\frac{2\ln r}{1-r}\biggr) \;, \nonumber\\
&& B(r)=\frac{1}{2(r-1)^2}\biggl(1+r+\frac{2r\ln r}{1-r}\biggr)\; ,\nonumber\\
&& C(r)=\frac{1}{6(r-1)^2}\biggl(10r-26+\frac{2r\ln r}{1-r}-\frac{18\ln r}{1-r}\biggr)\;.
\end{eqnarray}

The chromoelectric contributions to the quark EDMs are given by
\begin{eqnarray}
&& d_q^{\tilde g \; (C)}=\frac{g_s\alpha_s}{4\pi} \sum_{k=1}^{2}
     {\rm Im}(\Gamma_{q}^{1k}) \frac{M_3}{M_{\tilde{q}_k}^2}\;
      {\rm C}\biggl(\frac{M_3^2}{M_{\tilde{q}_k}^2}\biggr)\;,\nonumber\\
&& d_q^{\chi^+ \; (C)}=\frac{-g^2 g_s}{16\pi^2}\sum_{k=1}^{2}\sum_{i=1}^{2}
      {\rm Im}(\Gamma_{qik})
               \frac{m_{\chi^+_i}}{M_{\tilde{q}_k}^2}\;
                {\rm B}\biggl(\frac{m_{\chi^+_i}^2}{M_{\tilde{q}_k}^2}\biggr)\;,\nonumber\\
&& d_q^{\chi^0 \; (C)}=\frac{g_s g^2}{16\pi^2}\sum_{k=1}^{2}\sum_{i=1}^{4}
       {\rm Im}(\eta_{qik})
               \frac{m_{\chi^0_i}}{M_{\tilde{q}_k}^2}\;
                {\rm B}\biggl(\frac{m_{\chi^0_i}^2}{M_{\tilde{q}_k}^2}\biggr)\;.
\end{eqnarray}

Finally, the contribution to the Weinberg operator \cite{Weinberg:1989dx} from
the two-loop gluino-top-stop and gluino-bottom-sbottom diagrams reads
\begin{equation}
d^G=-3\alpha_s m_t \biggl(\frac{g_s}{4\pi}\biggr)^3\;
{\rm Im} (\Gamma_{t}^{12})\;\frac{z_1-z_2}{(M_3)^3}\;
{\rm H}(z_1,z_2,z_t) +\;(t\rightarrow b)\;,
\end{equation}
where $z_{i}=\biggl(\frac{M_{\tilde{t}_i}}{M_3}\biggr)^2,
z_t=\biggl(\frac{m_t}{M_3}\biggr)^2$. The two-loop function $H(z_1,z_2,z_t)$
is given by \cite{Dai:1990xh}
\begin{equation}
H(z_1,z_2,z_t)={1\over2}\int_0^1 dx \int_0^1 du \int_0^1 dy\; x(1-x)u{N_1 N_2\over D^4}\;,
\end{equation}
where
\begin{eqnarray}
&& N_1=u(1-x)+z_t x(1-x)(1-u)-2ux[z_1y+z_2(1-y)]\;,\nonumber\\
&& N_2=(1-x)^2 (1-u)^2 +u^2-{1\over 9}x^2(1-u)^2 \;,\nonumber\\
&& D=u(1-x) +z_t x(1-x)(1-u)+ux[z_1y+z_2(1-y)]\;.
\end{eqnarray}
The numerical behaviour of this function was studied in \cite{Dai:1990xh}. We emphasize
that the b-quark contribution is significant and often exceeds the top one.

In addition to the Weinberg two-loop diagram, there is another (Barr-Zee)
two-loop contribution which originates from the CP-odd Higgs exchange \cite{chang}.
Its numerical effect is however negligible  \cite{AKL}.

Let us now discuss possible ways to avoid overproduction of EDMs in supersymmetric
and string-inspired models. There are four known scenarios allowing to suppress the
EDM contributions: small SUSY CP-phases, heavy sfermions, EDM cancellations, and
flavor-off-diagonal CP violation.\\

\hspace{-0.5cm}{\it \bf \underline{1-- Small SUSY CP-phases:}}\\

For a light (below 1 TeV) supersymmetric spectrum,
the flavor--independent SUSY CP phases have to be small
in order to satisfy the experimental EDM bounds~\cite{AKL}. 

\begin{figure}[ht]
\epsfig{figure=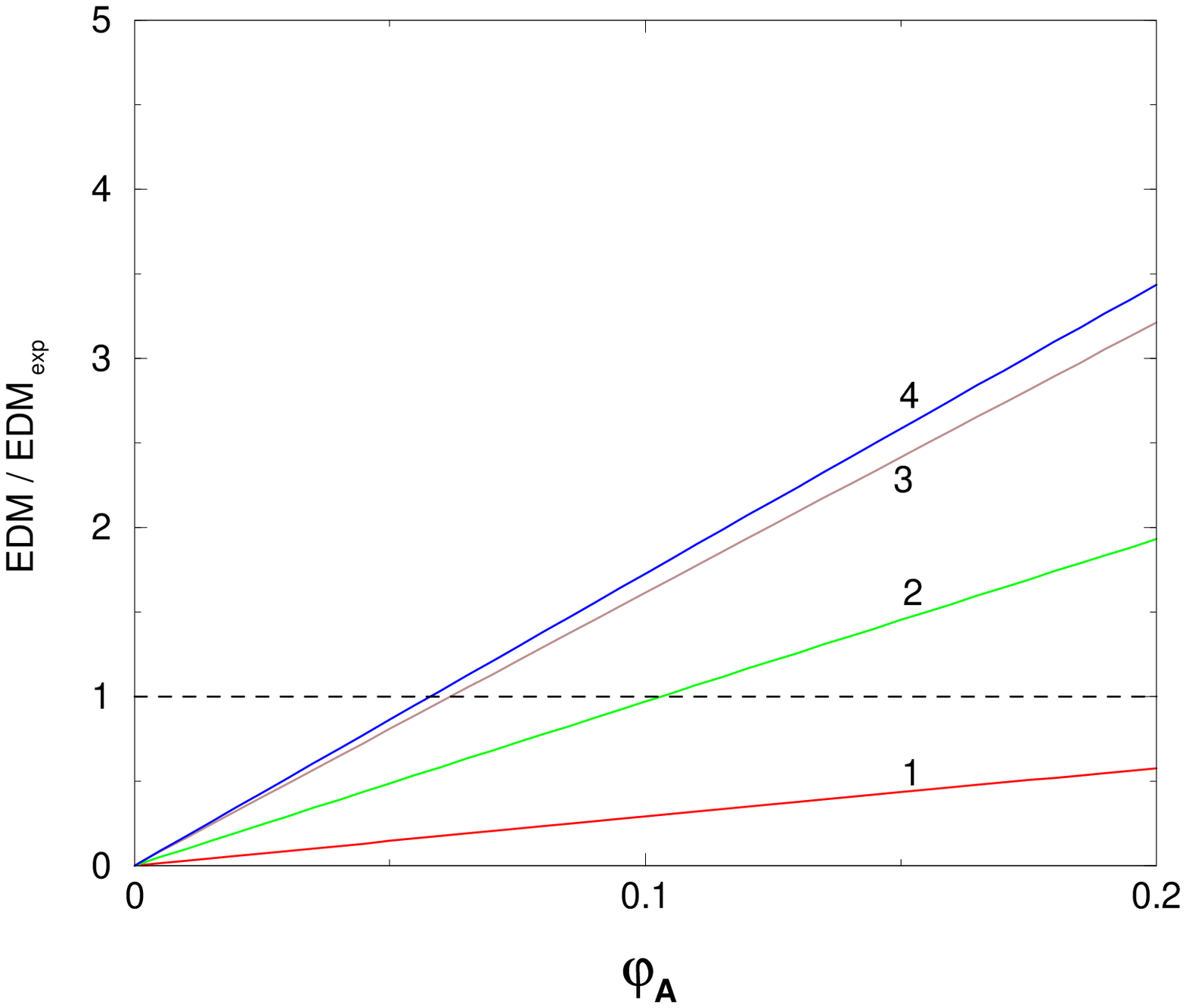,height=4.5cm,width=5.85cm,angle=0}
\hspace{1.cm}\epsfig{figure=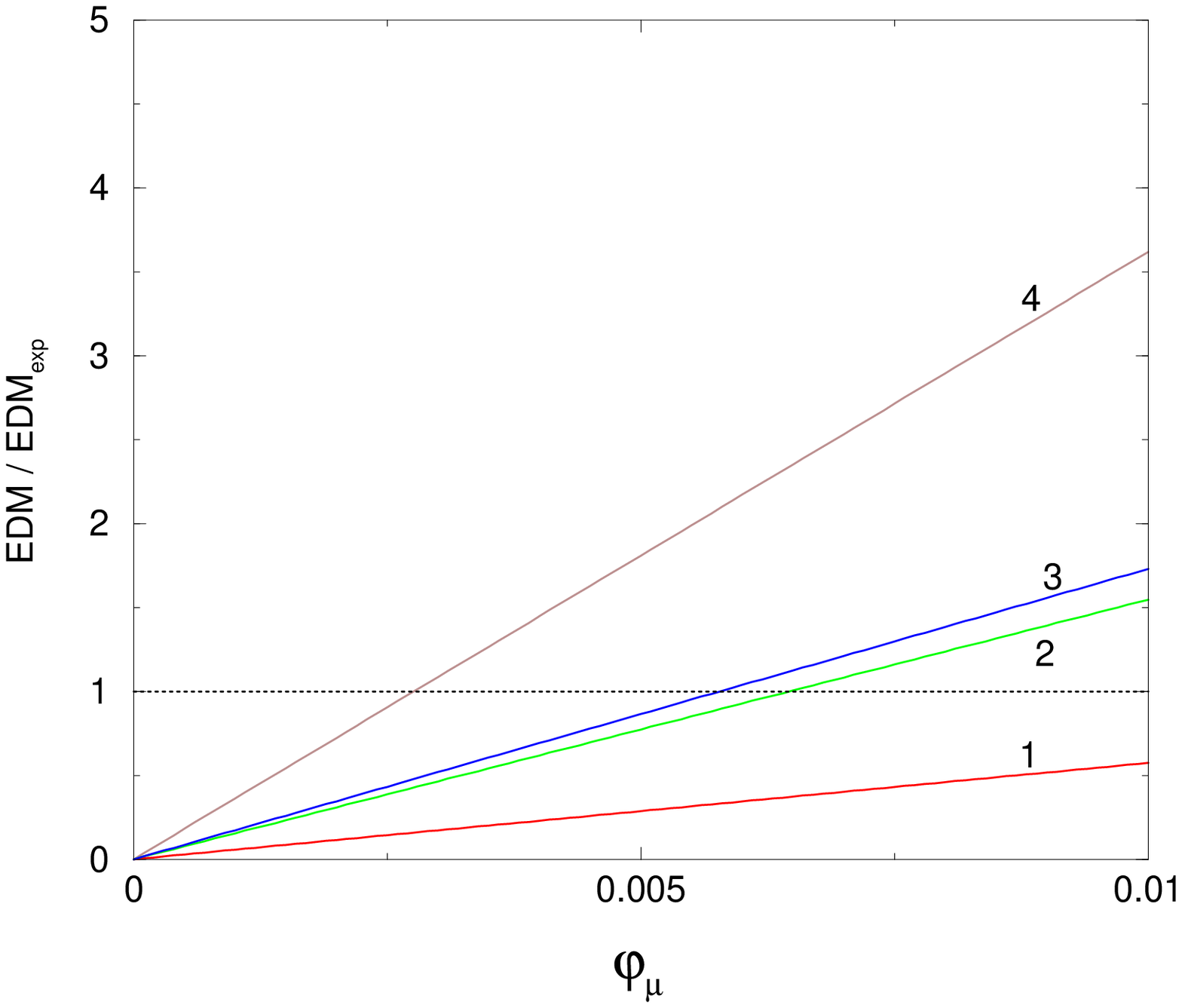,height=4.5cm,width=5.85cm,angle=0}
\medskip
\caption{EDMs as a function of $\phi_A$ (left) and $\phi_{\mu}$ (right).
 1 -- electron,  2 -- neutron (chiral model),
3 -- mercury, 4 -- neutron (parton model). 
The experimental limit is given by the horizontal line.
Here $\tan\beta=3$, $m_0=m_{1/2}=A=200$ GeV.}
\label{aphase}
\end{figure}

In Figs.\ref{aphase} we illustrate the EDMs
behaviour as a function of the CP-phases in the mSUGRA-type models, where 
we have set $m_0=m_{1/2}=A=200$ GeV.  
At low $\tan\beta$, the EDM constraints impose the following bounds 
(at the GUT scale):

\begin{eqnarray}
&& \phi_A \lsim 10^{-2}-10^{-1} \;, \nonumber\\
&& \phi_{\mu} \lsim 10^{-3}-10^{-2} \;, \nonumber\\
&& \phi_{M_i} \lsim 10^{-2} \;.
\label{smallph}
\end{eqnarray}
We note that $\phi_A$ is less constrained
than $\phi_{\mu}$ and $\phi_{M_i}$. There are two reasons for that:
first, $\phi_A$ is reduced by the RG running from the GUT scale down to the
electroweak scale and, second, the phase of the $(\delta_{11}^d)_{LR}$
mass insertion which gives the dominant contribution to the EDMs is more 
sensitive to $\phi_{\mu}$  and $\phi_{M_i}$ due to $\vert A \vert < \mu\tan\beta$.
We note that the bounds (\ref{smallph}) stay practically the same if we allow
for an ``uncertainty''  factor 2-3 in the overall EDM normalization.

Generally it is quite difficult to explain why the soft CP-phases have to be small.
In principle, small CP-phases could appear if CP were an approximate symmetry of
nature \cite{nir}.  However, recent experimental results show that CP violation
in the $B-\bar B$ mixing is large \cite{babar} and thus the approximate
CP hypothesis cannot be motivated.   \\

\hspace{-0.5cm}{\it \bf \underline{2-- Heavy sfermions:}}\\

This possibility is based on the decoupling of heavy supersymmetric particles.
Even if one allows ${\cal{O}}(1)$ CP violating phases, their effect will be
negligible if the SUSY spectrum is sufficiently heavy \cite{heavy}. Generally, SUSY fermions
are required to be lighter than the SUSY scalars by, for example, cosmological
considerations. So the decoupling scenario can be implemented with heavy
sfermions only.
Here the SUSY contributions to the EDMs are suppressed even
with maximal SUSY phases  because the squarks in the loop 
are very heavy and  the mixing angles are  small.
It was shown~\cite{AKL} that in order to satisfy the EDM of the mercury 
atom the sfermion masses have to be of order 10 TeV, which leads to a large 
hierarchy between SUSY and electroweak scales.
\begin{figure}[ht]
\begin{center}
\begin{tabular}{c}
\epsfig{file=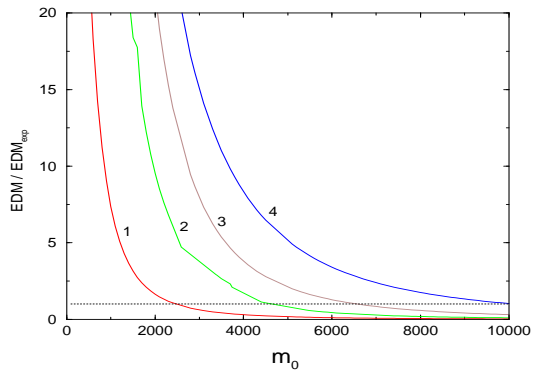, width=7cm, height=5cm}\\  
\end{tabular}
\end{center}
\caption{EDMs as a function of the universal mass parameter
$m_0$. 1 -- electron,  2 -- neutron (chiral model),
3 -- neutron (parton model), 4 -- mercury. 
The experimental limit is given by the horizontal line.
Here  $\tan\beta=3$, $m_{1/2}=A=200$ GeV, 
$\phi_{\mu}=\phi_A=\pi/2$.}
\label{heavy}
\end{figure}
In Fig.\ref{heavy} we display the EDMs as functions of the universal scalar 
mass parameter $m_0$ for the mSUGRA model with maximal CP-phases 
$\phi_{\mu}=\phi_A=\pi/2$ and $m_{1/2}=A=200$ GeV. 
We observe that all EDM constraints except for that of the electron require 
$m_0$ to be around 5 TeV or more. The mercury constraint is the strongest one and 
requires $m_0 \simeq 10\;{\rm TeV}$. 

This leads to a serious fine--tuning problem. Recall that one of the primary
motivations for supersymmetry was a solution to the naturalness problem. 
Certainly this motivation will be entirely lost if a SUSY model reintroduces
the same problem in a different sector, {\it i.e.,} for example a large
hierarchy between the scalar mass and the electroweak scale.
Also note that the dependence of the decoupling scale on the EDMs is quite slow. 
For instance, if we relax the mercury bound by a factor of 2 the decoupling 
scale changes from 10 TeV to about 8 TeV. Thus our conclusions are insensitive 
to the exact values of the EDMs.\\

\hspace{-0.5cm}{\it \bf \underline{3-- EDM cancellations:}}\\

The cancellation scenario is based on the fact that large cancellations among different
contributions to the EDMs are possible in certain regions of the parameter space 
\cite{nath,cancel,savoy,bartl} which allow for ${\cal{O}}(1)$ 
$flavour-independent$ CP phases. However, a recent analysis \cite{AKL}
has shown that this possibility is practically ruled out if in addition to 
the electron and neutron EDM constraints, the mercury constraint is imposed.

For the case of the electron, the EDM cancellations occur between the chargino and 
the neutralino contributions. For the case of the neutron and mercury, there are 
cancellations between the gluino and the chargino contributions as well as 
cancellations among contributions of different quarks to the total
EDM. It has been shown \cite{AKL} that the parameters allowing the EDM 
cancellations strongly depend on the neutron model.
For example, in the parton model, it is more difficult to achieve these cancellations
due to the large strange quark contribution. Therefore, one cannot restrict 
the parameter space in a model-independent way and caution is needed when 
dealing with the parameters allowed by the cancellations. 

\begin{figure}[ht]
\begin{center}
\begin{tabular}{c}
\epsfig{file=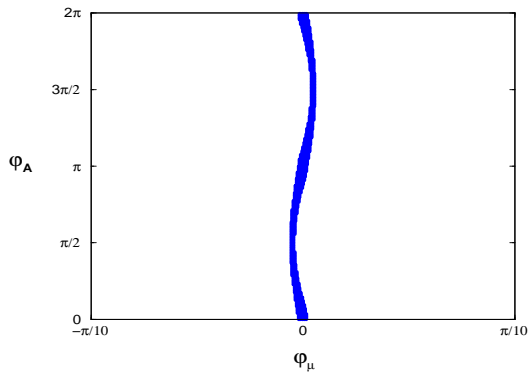, width=7cm, height=5cm}\\
\end{tabular}
\end{center}
\caption{Phases allowed by simultaneous electron, neutron, and mercury EDM
cancellations in mSUGRA. The $chiral$ quark neutron model is assumed.
Here  $\tan\beta=3$, $m_0=m_{1/2}=200$ GeV, $A=40$ GeV. }
\label{sugra}
\end{figure}
In mSUGRA, the EDM  cancellations can occur simultaneously for the electron, 
neutron, and mercury along a band in the $(\phi_A, \phi_{\mu})$ plane 
(Fig.\ref{sugra}). However, in this case the mercury constraint requires the 
$\mu$ phase to be  ${\cal{O}}(10^{-2})$ and the magnitude of the A-terms to be 
suppressed ($\sim 0.1m_0$) \cite{AKL} which results in only a small effect 
of the A-terms on the phase of the corresponding mass insertion. Thus the 
fact that the phase of the A-terms is 
unrestricted should not be attributed to the cancellations, but rather to the 
suppressed contribution of the A-terms. 

\begin{figure}[ht]
\begin{center}
\begin{tabular}{c}
\epsfig{file=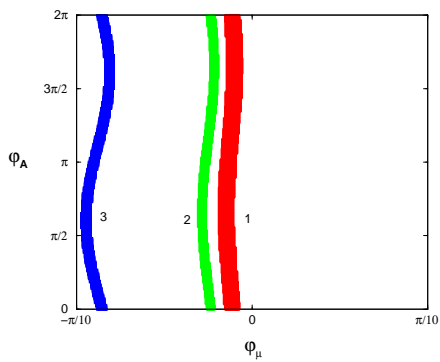, width=5.8cm, height=4.8cm}\\
\end{tabular}
\end{center}
\caption{ Bands allowed by  the electron (red), neutron (green), and mercury (blue) EDMs
cancellations in the mSUGRA-type model with  nonzero gluino and bino  phases.
Here  $\tan\beta=3$, $m_0=m_{1/2}=200$ GeV, $A=40$ GeV, $\phi_1=\phi_3=\pi/10$. }
\label{phi1phi3}
\end{figure}

If the gluino phase is turned on, simultaneous EEDM, NEDM, and mercury EDM  
cancellations are not possible. The gluino phase affects the NEDM cancellation 
band while leaving the EEDM cancellation band almost intact. 
An introduction of the bino phase $\phi_1$ qualitatively has the same 
``off-setting'' effect on the EEDM cancellation band as the gluino phase does on 
that of the NEDM .(Fig.\ref{phi1phi3}). Note that the bino phase has no 
significant effect on the 
neutron and mercury cancellation bands since the neutralino contribution in 
both cases is small. When both the gluino and bino phases are present 
(and fixed), simultaneous electron, neutron, and mercury EDMs cancellations
do not appear to be possible along a band. This conclusion has been also obtained
in type I string inspired models \cite{AKL}.

The cancellation scenario involves a significant fine-tuning. Indeed, restricting
the phases to the band where the cancellations occur does not increase the symmetry
of the model and thus is unnatural according to the t'Hooft's criterion 
\cite{hooft}.\\

\hspace{-0.5cm}{\it \bf \underline{4-- Flavor-off-diagonal CP violation:}}\\

Finally, we consider the possiblity that 
the SUSY CP violation have a flavor off-diagonal character just as in the 
SM~\cite{Abel:2000hn}. In the SM, CP violation appears  in flavor-changing processes 
which is one of the reasons why the predicted EDMs are small. A similar 
situation may occur in supersymmetric models.
In this case, the origin of the CP violation is 
assumed to be closely related to the origin of flavor structures 
rather than the origin of SUSY breaking. Thus
the flavor blind quantities as the $\mu$--term and gaugino masses
are real. As mentioned, the EDMs of the neutron, electron, and mercury atom 
impose severe constraints on the flavor-diagonal phases. 
Indeed, these are the only phases that enter into the EDM calculations.
However, this leaves the possibility that the flavor-off-diagonal phases 
are non-zero. Of course,  one has to specify the basis in which the considered 
quantity is classified as off-diagonal. 

This class of models requires hermitian Yukawa matrices and A-terms 
\cite{Abel:2000hn}, which forces the flavor-diagonal phases
to vanish (up to small RG corrections) in any basis. 
To see this, first note that the trilinear coupings
\begin{equation}
\hat A^\alpha_{ij}= A^\alpha_{ij}  Y^\alpha_{ij}
\end{equation}
are also hermitian. The quark Yukawa matrices are diagonalized by  
the $unitary$ transformation 
\begin{equation}
q \rightarrow V^q q ~\;\;,\;\;~
Y^q \rightarrow (V^q)^T ~ Y^q ~ (V^q)^* ~=~ {\rm diag}(h_1^q,h_2^q,h_3^q) \;\;, 
\end{equation}
such that the CKM matrix is $V_{CKM}= (V^u)^\dagger V^d$. If we transform 
the squark fields in the same manner, which defines the super-CKM basis, 
we will have
\begin{equation}
\hat A^q \rightarrow (V^q)^T ~ \hat A^q ~ (V^q)^* \;\;. 
\end{equation}
As a result, the trilinear couplings remain hermitian and the flavor-diagonal  
CP-phases inducing the EDMs vanish. Note that this argument would not work if 
in the original basis the trilinear couplings
but not the Yukawas were hermitian since the diagonalization would
require a biunitary transformation which would generally introduce the diagonal phases. 

The hermiticity is generally spoiled by the renormalization group (RG) running
from the high energy scale down to the electroweak scale (see, e.g. \cite{Bertolini:1991if}). 
In models under consideration we have the following setting at high energies:
\begin{eqnarray}
&& Y^{\alpha}=Y^{\alpha \dagger}\;, A^{\alpha}=A^{\alpha\dagger} \;, \nonumber\\
&& {\rm Arg} (M_k)= {\rm Arg} (B)    ={\rm Arg} (\mu)=0\;.
\end{eqnarray}
Generally, the off-diagonal elements of the A-terms can have ${\cal{O}}(1)$ 
phases without violating the EDM constraints. Due to the RG effects,  large 
phases in the soft trilinear couplings involving the third generation generate 
small phases in the flavour-diagonal mass insertions for the light generations,
and thus induce the EDMs. However, the amount of generated ``non-hermiticity'' 
depends on $\left[ Y^u,Y^d \right]$, $\left[ \hat A^u, \hat A^d \right]$, etc.
and thus is suppressed by the off-diagonal elements of the Yukawa matrices.
In particular, the size of the flavor-diagonal phases is quite sensitive to 
$Y_{13}$. For $Y_{13} \leq {\cal O}(10^{-3})$, the RG effects are unimportant 
and the induced EDMs are below the experimental limits.

In supergravity models, the trilinear parameters are given in 
terms of the K\"ahler potential and the Yukawa couplings 
\begin{eqnarray}
A_{\alpha \beta \gamma} = F^m \bigl[ \hat K_m  + \partial_m \ln
Y_{\alpha \beta \gamma} -\partial_m \ln (\tilde K_{\alpha} \tilde K_{\beta} 
\tilde K_{\gamma} ) \bigr] \;,
\label{A-terms}
\end{eqnarray}
where the Latin indices refer to the hidden sector fields that break 
SUSY and the Greek indices refer to the observable fields. 
According to our previous assumption, the $F^m$ is real. Also $\tilde 
K_{\alpha}$ and $\hat K_m$ are always real, thus the $A$--terms are 
Hermitian if the derivatives of the K\"ahler potential are either 
generation--independent or the same for the left and right fields of the 
same generation, \ie, if $\tilde{K}_{Q_{L_i}} \tilde{K}_{U_{R_j}}=
\tilde{K}_{Q_{L_j}} \tilde{K}_{U_{R_i}}$. 
These conditions are usually satisfied in string models.
It is interesting to note that although the SUSY breaking does not bring in
new source of CP violating, the trilinear soft parameters involve 
off--diagonal CP violating phases of $\mathcal{O}(1)$. This stems from the 
contribution of the term $\partial_m \ln Y_{\alpha\beta\gamma}$, which
has been found to be significant and sometimes even dominant in string 
models~\cite{stringcp}.
In what follows, we will show that these phases are unconstrained by the 
EDMs and will study their phenomenological implications in the $K$ and $B$ 
systems.

The relevant quantities appearing in the soft Lagrangian are $(Y^A_q)_{ij}=
(Y_q)_{ij} (A_q)_{ij}$ (indices not summed) which are also Hermitian at 
the GUT scale. For the sake of definiteness, we consider the following 
Hermitian Yukawa matrices at the GUT scale
\begin{eqnarray}\label{yukawa}
Y^u &= & \lambda_u \left(
\begin{array}{ccc}
5.94\times10^{-4} & 10^{-3}~i & -2.03\times10^{-2}  \\
- 10^{-3}~i & 5.07\times10^{-3} & 2.03\times10^{-5}~i \\
-2.03\times10^{-2} & -2.03\times10^{-5}~i & 1 
\end{array} \right) \;,\\
Y^d &=& \lambda_d \left(
\begin{array}{ccc}
6.84 \times 10^{-3} & (1.05+0.947~i) \times 10^{-2} & -0.023 \\
(1.05-0.947~i)\times 10^{-2} & 0.0489 & 0.0368~i\\
-0.023 & -0.0368~i & 1 
\end{array}\right),
\end{eqnarray}
where $\lambda_u=m_t/v \sin\beta$ and $\lambda_d=m_b/v\cos\beta$. 
These matrices reproduce, at low energy, the quark masses and the CKM
matrix. The renormalization group (RG) evolution of Yukawa couplings 
and the $A$ terms slightly violate the Hermiticity. Therefore, the resulting 
$Y^A_q$ at the electroweak scale has very small non--zero phases
in the diagonal elements (due to the large suppression from the 
off--diagonal entries of the Yukawa). 
However, what matters is the relevant phases appearing in the squark mass 
insertions in the super-CKM basis, \ie, the basis where the Yukawa matrices 
are diagonalized by a unitary transformation of the quark superfields 
$\hat{U}_{L,R}$ and $\hat{D}_{L,R}$ 
(Note that since the Yukawas are Hermitian matrices they are diagonalized
by one unitary transformation,\ie $V^q_L=V^q_R$).
Accordingly the trilinear terms $Y^A_q$ transform as $Y^A_q \to
V^{q^T} Y^A_q V^{q^*}$. Thus the $Y^A_q$ stay Hermitian to a very good degree
in the super-CKM basis. Therefore, the imaginary parts of the flavor conserving
mass insertions 
\begin{equation}
(\delta^{d(u)}_{ii})_{LR} = \frac{1}{m_{\tilde{q}}^2}
\left[ \left(V^{q^T} Y^A_q V^{q^*}\right)_{ii} v_{1(2)} -
\mu Y^{d(u)}_i v_2(1) \right],
\end{equation}
that appear in the EDM calculations are suppressed. In the above
formula the $m_{\tilde{q}}$ refers to the average squark mass and 
$v_i=\av{H_i^0}/\sqrt{2}$. The current experimental bound~\cite{AKL}
leads to the constraint $\mathrm{Im}(\delta^{d(u)}_{11})_{LR} 
\lsim 10^{-7}$. 

We start our analysis by revisiting the EDM constraints on the flavor
off diagonal phases of SUSY models with Hermitian Yukawa as in 
Eq.(\ref{yukawa}) and the following Hermitian $A$--terms\cite{Khalil:2002jq}:
\begin{equation}
A_d=A_u = \left( \begin{array}{ccc}
A_{11} & A_{12}~e^{i\varphi_{12}} & A_{13}~e^{i\varphi_{13}}  \\
A_{12}~e^{-i\varphi_{12}} & A_{22} &A_{23}~e^{i\varphi_{23}}  \\
A_{13}~e^{-i\varphi_{13}}&A_{23}~e^{-i\varphi_{23}}  & A_{33} 
\end{array} \right) \;.
\label{soft1}
\end{equation}
We also assume that the soft scalar masses and gaugino masses $M_a$
are given by
\bea
M_a &=& m_{1/2} ,~~ a=1,2,3,\\
m^2_Q &=&m_{H_1}^2 = m_{H_2}^2=m_0^2,\\
m^2_U &=& m^2_D=m_0^2~ \mathrm{diag}\{1,\delta_1,\delta_2\},
\label{soft2}
\eea
where the parameters $\delta_i$ and $A_{ij}$ can vary in the ranges
$[0,1]$ and $[-3,3]$ respectively. Note that in most string inspired 
models, the squark mass matrices are diagonal but not necessary
universal. The non--universality of the squark masses is not constrained
by the EDMs. However, this non--universality (specially between the 
first two generations of the squark doublets) is  severely constrained by 
$\Delta M_K$ and $\varepsilon_K$. 

\begin{figure}[t]
\begin{center}
\hspace*{-7mm}
\epsfig{file=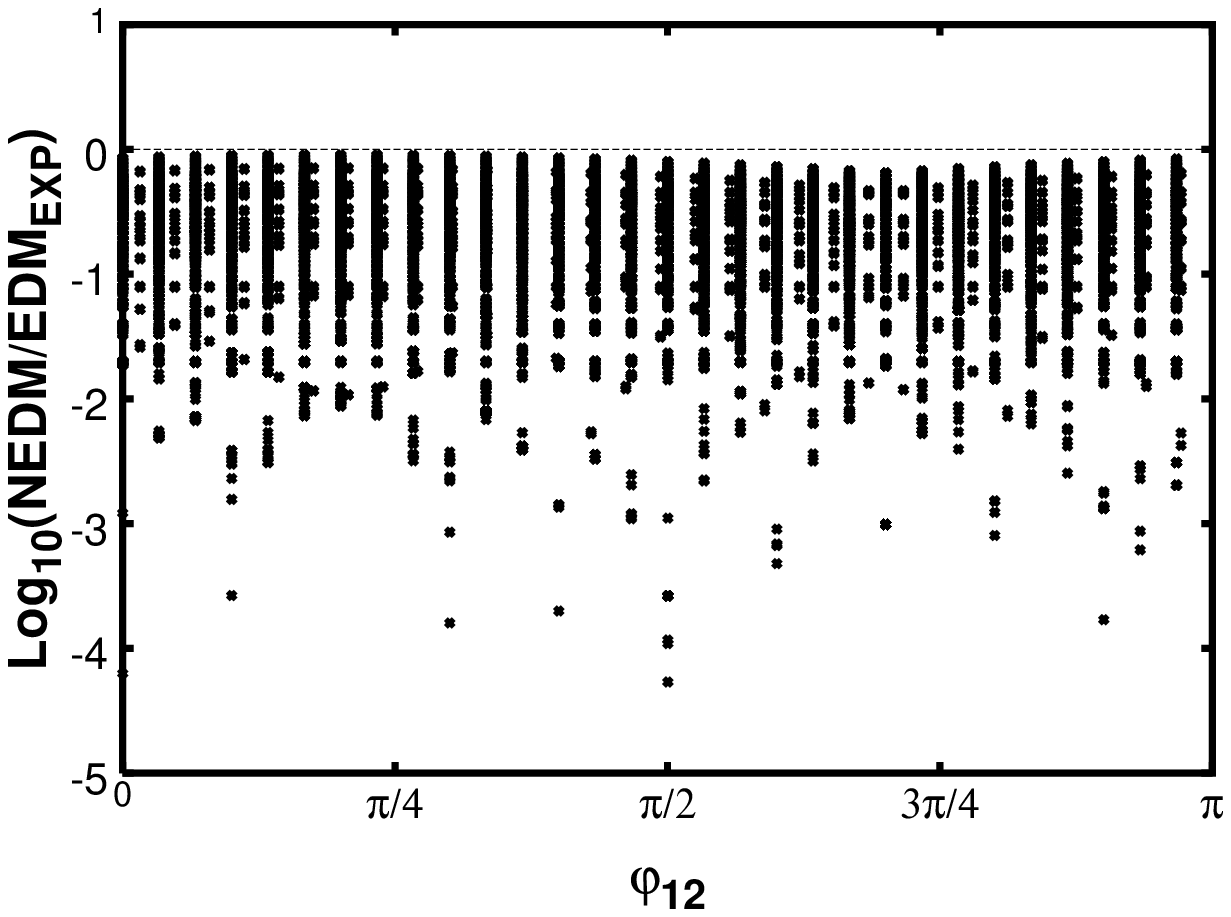,height=4.5cm,width=5.85cm} \quad
\epsfig{file=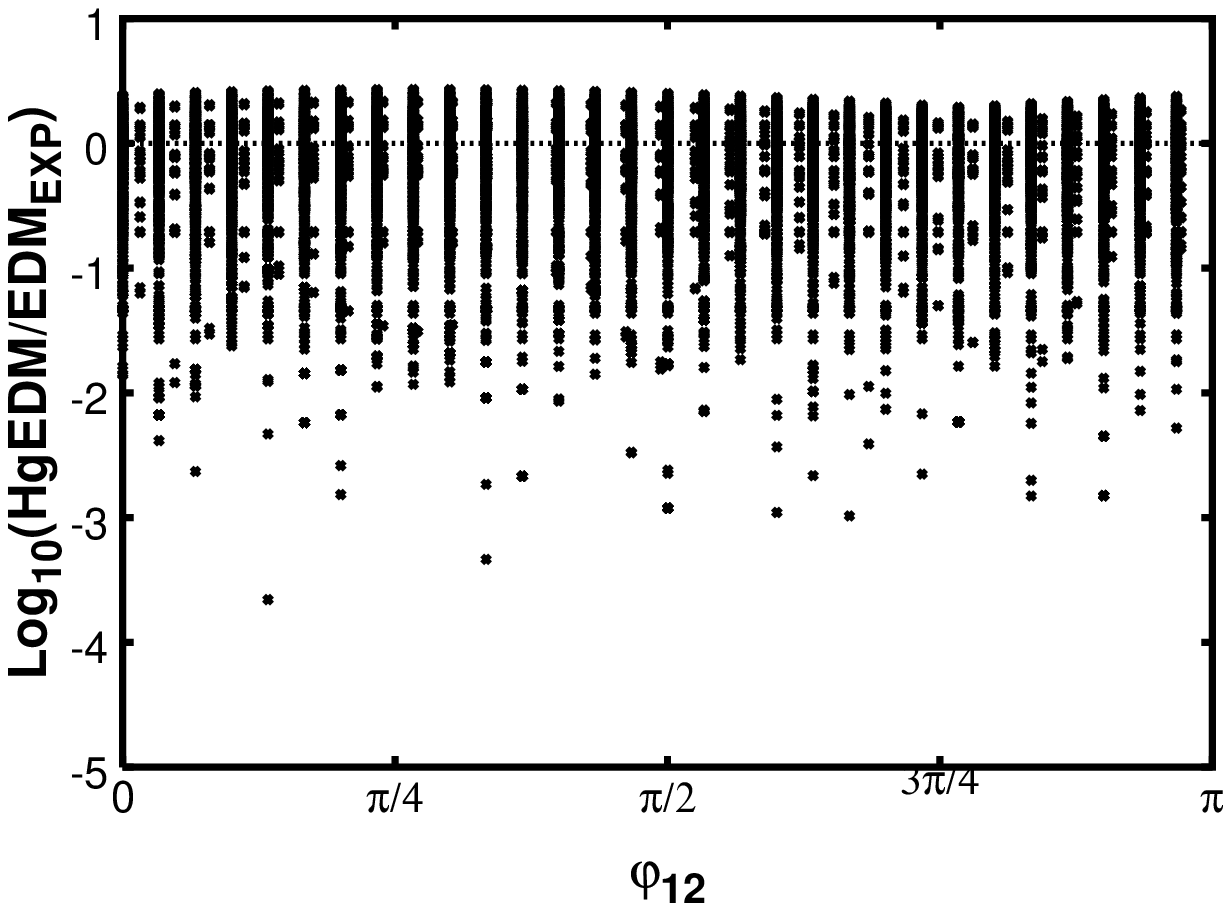,height=4.5cm,width=5.85cm}\\
\caption{{\small Neutron and mercury EDMs versus $\varphi_{12}$ 
for  $\tan\beta=5$ and $m_0=m_{1/2}=200$ GeV.}}
\label{edmherm}
\end{center}
\end{figure}

In Fig. \ref{edmherm} we display scatter plots for the neutron and mercury EDMs 
versus the phase $\varphi_{12}$ for $\tan \beta=5$, $m_0=m_{1/2}=200$ GeV,
$A_{ij}$ are scanned in the range $[-3,3]$, 
and the phases $\varphi_{13}$ and $\varphi_{23}$ are
randomly selected in the range $[0,\pi]$. As stated above, the parameters 
$\delta_i$ are irrelevant for the EDM calculations and we set them here 
to one. Finally, since $\mu$ is real the EDM results
display very little dependence on $\tan \beta$. 

It is important to mention that we have also imposed the constraints 
which come from the requirement of absence of charge and colour breaking 
minima as well as the requirement that the scalar potential be bounded
from below~\cite{CCB}. These conditions may be automatically
satisfied in minimal SUSY models, however in models with non--universal
$A$--terms they have to be explicitly checked. In fact, sometimes these
constraints are even stronger than the usual bounds set by the flavor
changing neutral currents~\cite{casas}.

As can be seen from Fig. \ref{edmherm}, the EDMs do not exceed the experimental 
bounds for most of the parameter space. Generally, they are one or two order of 
magnitude below the present limit, and the flavor--off diagonal phases of the 
$A$--terms can be $\mathcal{O}(1)$ without fine--tunning.
The points that lead to mercury EDM above the experimental bound 
correspond to $\varphi_{23}\simeq \pi/2$. This phase induces a considerable 
contribution to the chromoelectric EDM of the strange quark $C_2^s$. 
Thus the compatibility with mercury EDM experiment requires that the phase 
$\varphi_{23}$ should be slightly smaller than $\pi/2$.


\section{\bf \large CP violation in the Kaon system}
We have shown in the previous section that the Hermiticity of 
the Yukawa couplings and $A$--terms allows the existence of large 
off--diagonal SUSY CP violating phases while keeping the EDMs sufficiently 
small. However, the important question to address is whether these
``EDM--free'' phases can have any implication
on other CP violation experiments. In this section, we will 
concentrate on possible effects in the kaon system. 

%
\subsection{\bf \normalsize Indirect CP violation}

In the kaon system and due to a CP violation in $K^0-\bar K^0$ mixing, the 
neutral kaon mass eigenstates are superpositions of CP--even ($K_S$) and
CP--odd ($K_L$) components. However, the CP--odd $K_L$ decays into two pions 
through its small CP--even component. This decay, $K_L \to \pi \pi$, was the 
the first observation of CP violation. The measure for the indirect CP
violation is defined as 
\be
\varepsilon_K = {A(K_L\rightarrow \pi\pi)\over A(K_S\rightarrow
\pi\pi) }\;.
\ee
The experimental value for this parameter is 
$\varepsilon_K \simeq 2.28 \times 10^{-3}$.  Generally, 
$\varepsilon_K$ can be  calculated via
\be
\varepsilon_K = {1\over \sqrt{2} \Delta M_K} {\rm Im} \langle K^0
\vert H^{\Delta S=2}_{\rm eff} \vert \bar K^0 \rangle \;.
\label{deltamk}
\ee
Here $H^{\Delta S=2}_{\rm eff}$ is the effective Hamiltonian for
the $\Delta S=2$ transition. It can be expressed via the operator
product expansion as
\begin{equation}
H^{\Delta S=2}_{\rm eff}=\sum_{i} C_i(\mu) Q_i + h.c. \;.
\end{equation}
In the above formula, $C_i (\mu)$ are the Wilson coefficients 
and $Q_i$ are the EH local operators.\\
The relevant operators for the gluino contribution are~\cite{EH}
\bea\label{susy:eK:oper}
Q_1 & = & \bar{d}^{\alpha}_L \gamma_{\mu} s^{\alpha}_L \bar{d}^{\beta}_L 
\gamma^{\mu} s^{\beta}_L,~~~~~~
Q_2 = \bar{d}^{\alpha}_R s^{\alpha}_L \bar{d}^{\beta}_R 
s^{\beta}_L,~~~~~~~
Q_3  =  \bar{d}^{\alpha}_R s^{\beta}_L \bar{d}^{\beta}_R 
s^{\beta}_L,\nonumber\\
Q_4 &=&  \bar{d}^{\alpha}_R s^{\alpha}_L \bar{d}^{\beta}_L 
s^{\beta}_R,~~~~~~~~~~~~
Q_5 =  \bar{d}^{\alpha}_R s^{\beta}_L \bar{d}^{\beta}_L 
s^{\alpha}_R,
\eea
as well as the operators $\tilde{Q}_{1,2,3}$, that are 
obtained from $Q_{1,2,3}$ by the exchange $L \leftrightarrow R$. 
In the latter equations, $\alpha$ and $\beta$ are $SU(3)_c$ colour
indices, and the colour matrices obey the normalization 
$\Tr(t^a t^b)=\delta^{ab}/2$. Due to the gaugino dominance 
in the chargino--squark loop,
the most significant $\tilde{\chi}^\pm$ contribution is associated with the 
operator $Q_1$~\cite{olegchar}. 

In the presence of supersymmetric ($\tilde{g}$ and $\tilde{\chi}^\pm$) 
contributions, the 
following result for the amplitude $M_{12}$ is obtained:
\begin{equation}\label{susy:eK:fullm12}
M_{12} = M_{12}^{\mathrm{SM}} +  
M_{12}^{\tilde{g}} +  M_{12}^{\tilde{\chi}^\pm}\;.
\end{equation}
The SM contribution can be written as~\cite{buras:2001,branco:book}
\begin{equation}\label{susy:eK:m12sm}
\mathcal{M}_{12} = \frac{G_F^2 M^2_W}{12 \pi^2}
f^2_K m_K \hat{B}_K \mathcal{F}^* \;,
\end{equation}
where $\hat{B}_K$ is the renormalization group invariant which is 
given by $0.85\pm 0.15$, $f_K=160$~MeV is the K--meson decay constant, and 
$m_K= 490$ MeV the K--meson mass. The function $\mathcal{F}$ can be 
give as 
\begin{equation}\label{usy:cp:defF}
\mathcal{F} = \eta_1 \lambda_c^2 S_0(x_c) + \eta_2 \lambda_t^2
S_0(x_t) + 2 \eta_3 \lambda_c \lambda_t S_0(x_c,x_t)\;,
\end{equation}
and $\lambda_i=V_{is}^* V_{id}$, $\eta_i$ are
QCD correction factors ($\eta_1=1.38\pm0.20$, $\eta_2=0.57\pm0.01$,
$\eta_3=0.47\pm0.04$ - NLO values), and $S_0(x_i)$ are the loop functions given by
\bea
S_0(x_c)&=&x_c \;, \quad
S_0(x_t)=2.46{\left(\frac{m_t}{170\GeV}\right)}^{1.52} \nonumber\\
S_0(x_c,x_t)&=&x_c \left( \log \frac{x_t}{x_c} -
\frac{3x_t}{4(1-x_t)} - \frac{3 x_t^2 \log x_t}{4{(1-x_t)}^2} \right)
\eea

The supersymmetric term $M_{12}^{\tilde{g}}$ is given by
\bea\label{susy:eK:m12gluino}
M_{12}^{\tilde{g}} &=& \frac{-\alpha_S}{216 m_{\tilde{q}^2}} \frac{1}{3}
m_{K}f_{K}^{2}
\Biggl\{  \left( (\delta^d_{12})^2_{LL} +(\delta^d_{12})^2_{RR} \right) 
        B_1(\mu) \left(  24\,x\,f_6(x) + 66\,\tilde{f}_6(x) \right)
        \nonumber \\
&+&(\delta^d_{12})_{LL} (\delta^d_{12})_{RR} \left(\frac{M_{K}}{m_{s}
(\mu)+m_{d}(\mu)}\right)^{2}
\biggl[B_4(\mu) \left(378 \,x\,f_6(x) -54 \tilde{f}_6(x)\right) \nonumber\\
&+& B_5(\mu) \left(6 \,x\,f_6(x) +30 \tilde{f}_6(x)\right) \biggr]
+\left((\delta^d_{12})^2_{LR} + (\delta^d_{12})^2_{RL} \right)
\left(\frac{M_{K}}{m_{s}(\mu)+m_{d}(\mu)}\right)^{2}\nonumber\\
&&\left(-\frac{255}{2} 
B_2(\mu) -\frac{9}{2} B_3(\mu) \right) x\, f_6(x) +
(\delta^d_{12})_{LR} 
(\delta^d_{12})_{RL} \left(\frac{M_{K}}{m_{s}(\mu)+m_{d}(\mu)}
\right)^{2}\nonumber\\
&&\biggl[-99 B_4(\mu) - 45 B_5(\mu) \biggr]\tilde{f}_6(x)
\Biggr\},
\end{eqnarray}
where $x=(m_{\tilde{g}}/m_{\tilde{q}})^2$ and 
the functions $f_6(x),
\tilde{f}_6(x)$ are given by
\begin{eqnarray}
f_6(x) &=& \frac{6(1+3 x) \ln x + x^3 - 9 x^2 - 9 x + 17}{6(x-1)^5},\\
\tilde{f}_6(x) &=& \frac{6(1+3 x) \ln x + x^3 - 9 x^2 - 9 x + 17}{6(x-1)^5}.
\end{eqnarray}
The matrix elements of the operators $Q_i$ between the  $K$-meson states
in the vacuum insertion approximation (VIA), where $B=1$, can be found in 
Ref.\cite{Gabbiani}. The VIA 
generally gives only a rough estimate, so other methods, e.g. lattice QCD, 
are required to obtain a more realistic value. The matrix elements
of the renormalized operators can be written as~\cite{ciuchini}
\begin{eqnarray}
\langle \bar{K}^0 \vert Q_1(\mu) \vert K^0 \rangle &=& \frac{1}{3} M_K
f_K^2 B_1(\mu),\\
\langle \bar{K}^0 \vert Q_2(\mu) \vert K^0 \rangle &=& -\frac{5}{24}
\left(\frac{M_K}{m_s(\mu) + m_d (\mu)}\right)^2 M_K
f_K^2 B_2(\mu),\\
\langle \bar{K}^0 \vert Q_3 (\mu) \vert K^0 \rangle &=& -\frac{1}{24}
\left(\frac{M_K}{m_s(\mu) + m_d (\mu)}\right)^2 M_K
f_K^2 B_3(\mu),\\
\langle \bar{K}^0 \vert Q_4(\mu) \vert K^0 \rangle &=& -\frac{5}{24}
\left(\frac{M_K}{m_s(\mu) + m_d (\mu)}\right)^2 M_K
f_K^2 B_4(\mu)\;,
\end{eqnarray}
where $Q_i(\mu)$ are the operators renormalized at the scale $\mu$.
The expressions of the matrix elements of the operators $Q_{1-3}$ 
are valid for the operators $\tilde{Q}_{1-3}$ \cite{ciuchini}, and for 
$\mu =2$ GeV we have \cite{ciuchini}
\bea\label{susy:eK:Bpar}
B_1(\mu) &=& 0.60\;, ~~~~~~~~~~ B_2(\mu) = 0.66\;, ~~~~~~~~~~ 
B_3(\mu) = 1.05\;,\nonumber\\
B_4(\mu) &=& 1.03\;,~~~~~~~~~~ B_5(\mu) = 0.73\;. \nonumber
\eea
Using these values, the gluino contribution to the $K^0-\bar K^0$ can be 
calculated
via Eq.(\ref{deltamk}). As mentioned above,
for universal soft scalar masses the LL and RR insertions are generated 
only through the RG running and can be neglected. 
The LR and RL mass insertions appear at the tree level and may have 
tangible effects. It is worth mentioning that, the RL and LR mass insertions
contribute with the same sign in Eq.(\ref{susy:eK:m12gluino}) and for 
Hermitian A-terms
they are almost equal, so no cancellation between these two contributions 
occurs. 

\begin{figure}[t]
\begin{center}
\hspace*{-7mm}
\epsfig{file=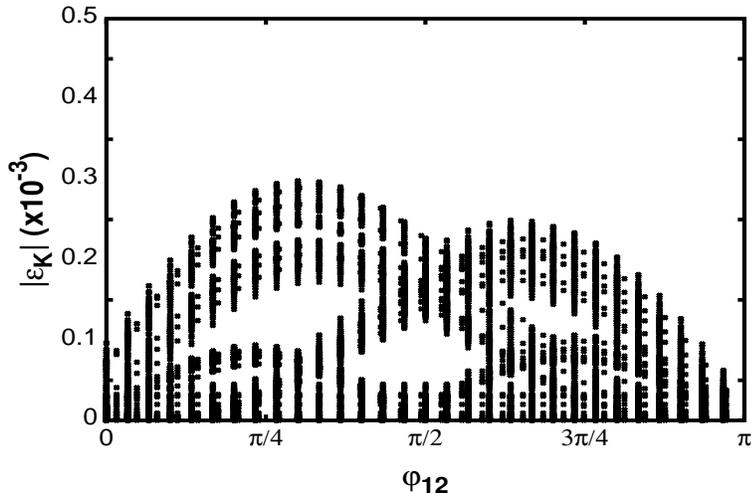,width=10cm,height=6.25cm}\\
\caption{{\small The gluino contribution to $\vert \varepsilon_K \vert$
as a function of $\varphi_{12}$ for 
$\delta_i=1$,  $\tan\beta=5$, and $m_0=m_{1/2}=200$ GeV.}}
\label{fig7}
\end{center}
\end{figure}

In Fig. \ref{fig7} we plot the values of $\vert \varepsilon_K \vert$ versus 
the phase $\varphi_{12}$ for $\delta_i=1$ and the other parameters are chosen 
as in Fig. \ref{edmherm}. From this figure, we conclude 
that the SUSY contribution with Hermitian Yukawa and universal soft scalar
masses can not account for the experimentally observed indirect CP
violation in the Kaon system. In Ref.\cite{Abel:2000hnl} values for
$\varepsilon_K\sim 10^{-3}$ have been obtained but these values require 
light gaugino mass ($m_{1/2}\sim 100$ GeV) which is now excluded by the new
experimental limits on the mass of the lightets Higgs. Also it requires
that the magnitude of the off--diagonal entries of the $A$--terms should be
much larger (at least five times larger) than the diagonal ones, which
looks unnatural. 

The chargino contribution to the $K^0-\bar{K}^0$ mixing is given 
by~\cite{olegchar}
\begin{equation}\label{susy:eK:m12char}
M_{12}^{\tilde{\chi}^\pm} = \frac{g^2}{768 \pi^2 m_{\tilde{q}^2}} \frac{1}{3}
m_{K}f_{K}^{2} B_1(\mu) \left(\sum_{a,b} K_{a2}^* 
(\delta^u_{LL})_{ab} K_{b1}\right)^2
\sum_{i,j} \vert V_{i1}\vert^2 \vert V_{j1} \vert^2 H(x_i,x_j),
\end{equation}
where $x_i=(m_{\tilde{\chi}_i^+}/m_{\tilde{q}})^2$. 
Here $K$ is the CKM matrix, 
$a,b$ are flavour indices, $i,j$ label the 
chargino mass eigenstates, and $V$ is one of the matrices used for
diagonalizing the chargino mass matrix.
The loop function $H(x_i, x_j)$ is given in Ref.\cite{olegchar}.
However, the chargino contribution can be significant only if there is a 
large LL mixing in the up- sector, namely  Im$(\delta^u_{LL})_{21} 
\sim 10^{-3}$ and Re$(\delta^u_{LL})_{21} \sim 10^{-2}$~\cite{olegchar}.
Such mixing can not be accommodated with the universal scalar masses assumption
(\ie, $\delta_i=1$). In this case, the values of the 
$\mathrm{Im}(\delta^u_{LL})$ are of order $10^{-6}$ which leads to a 
negligible chargino contribution to $\varepsilon_K$.
With non--universal soft scalar masses ($\delta_i \neq 1$) a possible 
enhancement in the chargino contribution is expected however, as we will 
show, this non--universality also leads to a larger enhancement in the 
gluino contribution.
So, the dominant SUSY contribution to the $K-\bar K$ mixing in 
this class of models will be provided by the gluino exchange diagrams.

A possible way to enhance $\varepsilon_K$ is to have non--universal
soft squark masses at GUT scale~\cite{Khalil:2002jq}. As mentioned above, the soft scalar masses
are not necessarily universal in generic SUSY models and their 
non--universality is not constrained by the EDMs. Thus for 
$\delta_i \neq 1$ the mass insertion $(\delta^d_{12})_{RR}$ is enhanced and 
we can easily saturate $\varepsilon_K$ through the gluino contribution.
To see this more explicitly, let us consider the LL and RR squark mass matrices
in the super--CKM basis
\bea
\left(M^2_{\tilde{d}}\right)_{LL}&\sim &V^{d\dagger}~ M^2_{Q}~
V^{d },\nonumber\\
\left(M^2_{\tilde{d}}\right)_{RR}&\sim&V^{d\dagger}~(M^2_{D})^T~ 
V^{d }\;.
\eea
Due to the universality assumption of $M_Q^2$ at GUT scale, the matrix
$\left(M^2_{\tilde{d}}\right)_{LL}$ remains approximately universal and 
the mass insertions $(\delta_{12}^d)_{LL}$ are sufficiently small 
($\mathrm{Im}(\delta_{12}^d)_{LL} \sim 10^{-5}$). However, since the masses
of the squark singlets $M^2_D$ are not universal, Eq.(\ref{soft2}),
sizeable off--diagonal elements in $\left(M^2_{\tilde{d}}\right)_{RR}$
are obtained. We find that $\mathrm{Re}(\delta_{12}^d)_{RR}$ is enhanced
from $\sim 10^{-7}$ in the universal case ($\delta_i=1$) to 
$\sim 10^{-3}$ for $\delta_i\sim 0.7$ while the 
imagenary part remains the same, of order $10^{-7}$. Thus, in this case, 
we have 
$$\sqrt{\vert \mathrm{Im}\left((\delta^d_{12})_{LL} (\delta^d_{12})_{RR}
\right)\vert} \simeq \sqrt{\vert \mathrm{Re} (\delta^d_{12})_{RR}
\mathrm{Im}(\delta^d_{12})_{LL}\vert}\simeq 10^{-4}$$ 
which is the required value in order to saturate the observed result of 
$\varepsilon_K$~\cite{Gabbiani}.  

\begin{figure}[t]
\begin{center}
\hspace*{-7mm}
\epsfig{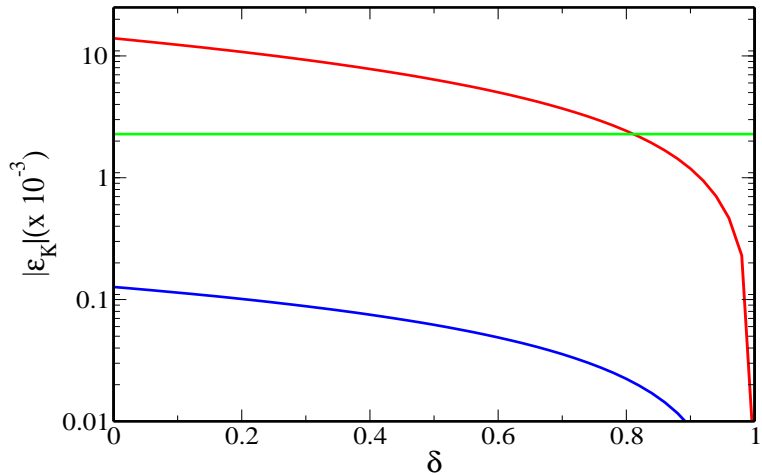}\\
\caption{{\small The value of $\vert \varepsilon_K \vert$
as a function of the parameters $\delta_1$ (upper curve) and
$\delta_2$ (lower curve) for $\tan\beta=5$, and $m_0=m_{1/2}=200$ GeV.}}
\label{fig8}
\end{center}
\end{figure}

In Fig. \ref{fig8} we show the dependence of $\vert \varepsilon_K \vert$
on the parameters $\delta_i$. There, the two curves, from top to bottom, 
correspond to the values of $\vert \varepsilon_K \vert$ versus 
$\delta_1$ (for $\delta_2=1$) and $\delta_2$ (for $\delta_1=1$) respectively. 
As explained above, in this scenario the main contribution of 
$\varepsilon_K$ is due to 
LL and RR sectors and the LR sector has essentially no effect on  
$\varepsilon_K$. We also see that any non--universality between
the soft scalar masses of the third and the first two generations
can not lead to significant contribution to $\varepsilon_K$ and 
some spliting between the scalar masses of the first two generations
is necessary. This stems from the fact that the effect of the third 
generation on the mass insertion $(\delta_{12}^d)_{RR}$ is suppressed 
by $V_{13} \sim \mathcal{O}(10^{-2})$ while $V_{12}\sim \sin\theta_C$. 
Finally, as we can see from this figure, in order to avoid over saturation 
of the experimental value of $\varepsilon_K$, the parameter $\delta_1$ should 
be of order 0.8.

%
\subsection{\bf \normalsize Direct CP violation}

Next let us consider  SUSY contributions to $\varepsilon^{\prime}/\varepsilon$.
The ratio $\varepsilon^{\prime}/\varepsilon$ is a measure of 
direct CP violation in the $K \to \pi \pi$ decays and is given by
\begin{equation}
\varepsilon'/\varepsilon=-{\omega \over \sqrt{2}~\vert \varepsilon 
\vert~ {\rm Re}A_0}~ \left( {\rm Im} A_0 -{1\over \omega}~{\rm Im}A_2 \right) ,
\label{eprimeformula}
\end{equation}
where $A_{0,2}$ are the amplitudes for the $\Delta I=1/2,3/2$ transitions, 
and $\omega\equiv {\rm Re} A_2 /{\rm Re} A_0 \simeq 1/22 $ reflects
the strong enhancement of $\Delta I= 1/2$ transitions over those with
$\Delta I= 3/2$. Experimentally it has been found to be 
${\rm Re}(\varepsilon'/\varepsilon) \simeq 1.9 \times 10^{-3}$ 
which provides firm evidence for the existence of direct CP violation. 
The $\mathrm{Im} A_{0,2}$ are calculated from the general low energy
effective Hamiltonian for $\Delta S=1$ transition,
\begin{eqnarray}
&& H_{\rm eff}^{\Delta S=1}= \sum_i C_i(\mu)  O_i +h.c. \;,
\end{eqnarray}
where $C_i$ are the Wilson coefficients and the list of the relevant 
operators $O_i$ for this transition is given in 
Ref.\cite{buras2,emidio,buras3}. Let us recall here that these 
operators can be classified into three categories. The first category 
includes dimension six operators: $O_{1,2}$ which refer to 
the current-current operators, $O_{3-6}$ for QCD penguin operators and 
$O_{7-10}$ for electroweak penguin operators~\cite{buras2}. 
The second category includes dimension five operators: magnetic- 
and electric-dipole penguin operators $O_{g}$ and $O_{\gamma}$ which are 
induced by the gluino exchange~\cite{emidio}. 
The third category includes the only dimension 
four operator $O_Z$ generated by the $\bar{s}dZ$ vertex which is mediated 
by chargino exchanges~\cite{buras3}. In addition, one should take 
into account $\tilde{O}_i$ operators which are obtained from $O_i$ 
by the exchange $L \leftrightarrow R$.

In spite of the presence of this large number of operators that
in principle can contribute to $\varepsilon'/\varepsilon$, it is 
remarkable that few of them can give significant contributions.
As we will discuss below, this is due to the suppression of the matrix
elements and/or the associated Wilson coefficients of most of the operators.
The SM contribution to $\varepsilon'/\varepsilon$ is dominated by the 
operators $Q_6$ and $Q_8$, and can be expressed
as
\begin{equation}
\mathrm{Re}\left(\frac{\varepsilon^\prime}{\varepsilon}\right)^{\mathrm{SM}} = 
\frac{\mathrm{Im}\left(\lambda_t \lambda_u^*\right)}{\vert \lambda_u \vert}~ 
F_{\varepsilon^\prime},
\end{equation}
where $\lambda_i= V_{is}^* V_{id}$ and the function 
$F_{\varepsilon^\prime}$ is given in Ref.\cite{buras:2001}.
By using our Hermitian Yukawa in Eq.(\ref{yukawa}) we
get 
\be 
\varepsilon'/\varepsilon \simeq 7.5 \times 10^{-4}.
\ee
Again, the SM prediction is below the observed value. Nevertheless,
this estimat can not be considered as a firm conclusion for 
a new physics beyond the SM since there are significant hadronic 
uncertainties are involved.

The supersymmetric contribution to $\varepsilon'/\varepsilon$ depends
on the flavor structure of the SUSY model. It is known that, in a minimal 
flavor SUSY model, it is not possible to generate a sizeable contribution
to $\varepsilon'/\varepsilon$ even if the SUSY phases are assumed to be large. 
Recently, it has been pointed out that with non--degenerate $A$--terms 
the gluino contribution to $\varepsilon'/\varepsilon$ can naturally be 
enhanced to saturate the observed value~\cite{susyK}. 
Indeed, in this scenario, 
the LR mass insertions can have large imaginary parts and the chromomagnetic 
operator $O_g$ gives the dominant contribution to $\varepsilon'/\varepsilon$
\begin{equation}
\mathrm{Re}\left(\frac{\varepsilon^\prime}{\varepsilon}
\right)^g \simeq 
\frac{11 \sqrt{3}}{64 \pi^2 \vert \varepsilon \vert 
\mathrm{Re} A_0}~  \frac{m_s}{m_s+m_d} \frac{F_k^2}{F_{\pi}^3}~ m_K^2~
m_{\pi}^2~ \mathrm{Im}\left[ C_g - \tilde{C}_g\right] \;,
\end{equation}
where $C_g$ is the Wilson coefficient associated with the operator $O_g$,
given by
\be
C_{g} = \frac{\alpha_s \pi}{m_{\tilde{q}}^2} \left[
(\delta^d_{12})_{LL} 
\left(-\frac{1}{3} M_3(x) - 3 M_4(x) \right) +
(\delta^d_{12})_{LR} \frac{m_{\tilde{g}}}{m_s} 
\left(-\frac{1}{3} M_1(x) - 3 M_2(x) \right)
\right] \;,
\ee
where the functions $M_i(x)$ can be found in Ref.\cite{Gabbiani} and  
$x=m_{\tilde{g}}^2/m_{\tilde q}^2$.

Using these relations, one finds that in order to saturate 
$\varepsilon'/\varepsilon$ from the gluino contribution
the imaginary parts of the LR mass insertions for $x\simeq 1$ should satisfy
$\mathrm{Im}(\delta^d_{12})_{LR} \sim 10^{-5}$. Such values can easily
be obtained in this class of models. However, as mentioned above, in the 
case of Hermitian $A$--terms and Yukawa couplings we have 
$(\delta^d_{12})_{LR} \simeq (\delta^d_{12})_{RL}$, hence we get
$\mathrm{Im}\left[ C_g - \tilde{C}_g\right]\simeq \mathrm{Im}
\left[(\delta^d_{12})_{LR} - (\delta^d_{12})_{RL} \right] \simeq 10^{-6}$
which leads to a negligible gluino contribution to $\varepsilon'/\varepsilon$
\cite{Abel:2000hn}. It is worth noticing that, due to the universality assumption
of $M_{\tilde{Q}_L}^2$, the imaginary part of the mass insertion 
$(\delta^d_{12})_{LL}$ is of order $10^{-5}$. So its contribution to 
$C_g$ is negligible with respect to the LR one which is enhanced by the ratio
$m_{\tilde{g}}/m_s$. To achieve the required contribution to 
$\varepsilon'/\varepsilon$ from the LL sector, one has to relax this 
universality assumption\cite{Khalil:2002jq} to get $\mathrm{Im}(\delta_{12}^d)_{LL}\sim 10^{-2}$. 
However, as we will discuss below, in this case the chargino contribution 
is also enhanced and becomes dominant.

Now we turn to the chargino contributions. The dominant contribution 
is found to be due to the terms proportional to a single 
mass insertion~\cite{olegchar}.
\begin{equation}\label{susy:char}
\mathrm{Re}\left(\frac{\varepsilon^\prime}{\varepsilon}\right)^{\chi^{\pm}}=
\mathrm{Im} \left(\sum_{a,b} K_{a2}^*
(\delta^u_{ab})_{LL} K_{b1}\right) \; F_{\varepsilon^\prime}(x_{q\chi})\;.
\end{equation}
The function $F_{\varepsilon^\prime}(x_{q\chi})$,
where $x_{q\chi}=m_{\tilde{\chi}^\pm}^2/m_{\tilde q}^2$,
is given in~\cite{olegchar}.
We find that the contributions involving a double mass insertion, 
like those arising from the supersymmetric effective $\bar{s}dZ$, can not give
any significant contribution, however we take them into account.
The above contribution is dominated by 
$(\delta^u_{12})_{LL}$ and in order to account for $\varepsilon'/\varepsilon$
entirely from the chargino exchange the up sector has to employ
a large LL mixing. Again, with universal $M_{\tilde{Q}}^2$,  
$\mathrm{Im}(\delta^u_{12})_{LL}\sim 10^{-6}$ and the chargino 
contributions (as the gluino one) to $\varepsilon'/\varepsilon$ is 
negligible.  

Finally, we consider another possibility proposed by Kagan and Neubert 
to obtain a large contribution to $\varepsilon'/\varepsilon$ \cite{neubert}.
It is important to note that in the previous mechanisms to generate 
$\varepsilon'/\varepsilon$ one is tacitly assuming that $\Delta I=1/2$
transitions are dominant and that the $\Delta I=3/2$ ones are suppressed
as in the SM. However, in Ref.\cite{neubert} it was shown that it is
possible to generate a large $\varepsilon'/\varepsilon$ from the
$\Delta I=3/2$ penguin operators. This mechanism relies on the LL
mass insertion $(\delta_{21}^d)_{LL}$ and requires isospin violation
in the squark masses $(m_{\tilde{u}} \neq m_{\tilde{d}})$. 
In this case, the relevant $\Delta S=1$ gluino box diagrams lead to 
the effective Hamiltonian \cite{neubert}
\be
H_{\mathrm{eff}}= \frac{G_F}{\sqrt{2}} \sum_{i=3}^6 
\left( C_i(\mu) Q_i + \tilde{C}_i \tilde{Q}_i \right) + h.c. 
\ee
where 
\bea 
Q_1 &=& (\bar{d}_{\alpha} s_{\alpha})_{V-A}~
(\bar{q}_{\beta} q_{\beta})_{V+A},~~~~ 
Q_2 = (\bar{d}_{\alpha} s_{\beta})_{V-A}~
(\bar{q}_{\beta} s_{\alpha})_{V+A},\\
Q_3 &=& (\bar{d}_{\alpha} s_{\alpha})_{V-A}~
(\bar{q}_{\beta} q_{\beta})_{V-A},~~~~ 
Q_4 = (\bar{d}_{\alpha} s_{\beta})_{V-A}~
(\bar{q}_{\beta} q_{\alpha})_{V-A},
\eea
and the operators $\tilde{Q}_i$ are obtained from $Q_i$ by
exchanging $L \leftrightarrow R$. It turns out that the SUSY 
$\Delta I=3/2$ contribution to $\varepsilon'/\varepsilon$
is given by~\cite{neubert}
\be
\mathrm{Re}\left(\frac{\varepsilon^\prime}{\varepsilon}\right)^{\Delta I=3/2}
\simeq 19.2 \left(\frac{500 \mathrm{GeV}}{m_{\tilde{g}}}\right)^2 
B_8^{(2)} K(x_d^L, x_u^R, x_d^R)~ \mathrm{Im}(\delta^d_{12})_{LL} .
\ee
Here, $x_u^{L,R}= \left(\frac{m_{\tilde{u}_{L,R}}}{m_{\tilde{g}}}\right)^2$,
$x_d^{L,R}= \left(\frac{m_{\tilde{d}_{L,R}}}{m_{\tilde{g}}}\right)^2$,
$B_8^{(2)}(m_c)\simeq 1$ and the function $K(x,y,z)$ is given by
\be
K(x,y,z) = \frac{32}{27} \left[ f(x,y) - f(x,z) \right] 
+ \frac{2}{27}\left[ g(x,y) - g(x,z) \right],
\ee
where $f(x,y)$ and $g(x,y)$ are given in Ref.\cite{neubert}. 
It is clear from the above equation that for $m_{\tilde{d}_R}=
m_{\tilde{u}_R}$, the function $K(x_d^L, x_u^R, x_d^R)$ vanishes identically.
Thus, a mass splitting between the right--handed squark mass is 
necessary to get large contributions to $\varepsilon'/\varepsilon$ through 
this mechanism. Furthermore, the $\mathrm{Im}(\delta^d_{12})_{LL}$ has to 
be of order $\mathcal{O}(10^{-3} - 10^{-2})$ to saturate the observed 
value of $\varepsilon'/\varepsilon$. It is clear that, with universal 
$M^2_{\tilde{Q}}$, this contribution can not give any significant value
for $\varepsilon'/\varepsilon$. 

Now we relax the universality assumption of $M^2_{\tilde{Q}}$ at GUT scale
to enhance the mass insertion  $(\delta^d_{12})_{LL}$ and saturate the 
experimental value of $\varepsilon'/\varepsilon$ \cite{Khalil:2002jq}. As mentioned above, 
the non--universality between the first two generation of $M^2_{\tilde{Q}}$
is very constrained by $\Delta M_K$ and $\varepsilon_K$. Therefore we just 
assume that the mass of third generation is given by $\delta_3 m_0$ while
the masses of the first two generations remain universal and equal to
$m_0$. This deviation from universality provides enhancement to both
$\varepsilon_K$ and $\varepsilon'/\varepsilon$. We have chosen the parameters
$\delta_i$ so that the total contributions of $\varepsilon_K$ from chargino 
and gluino are  consistent with the experimental limits.

\begin{figure}[t]
\begin{center}
\hspace*{-7mm}
\epsfig{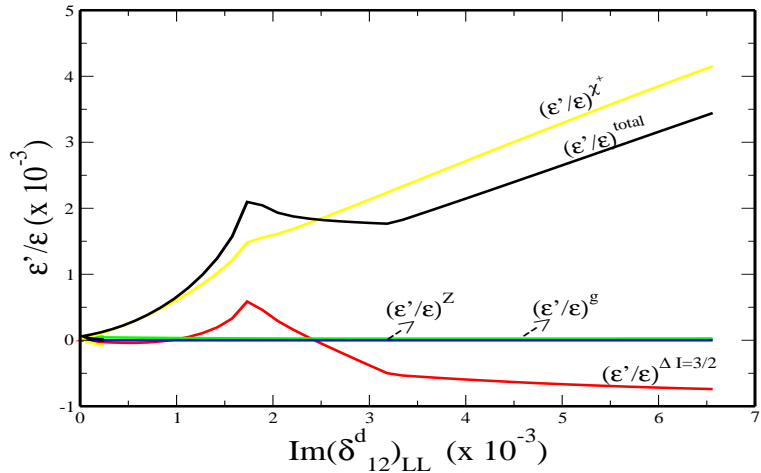}\\
\caption{{\small The  $\varepsilon'/\varepsilon$ contributions 
versus the Imaginary part of the mass insertion $(\delta^d_{12})_{LL}$.}}
\label{fig9}
\end{center}
\end{figure}

In Fig. \ref{fig9} we present the different gluino and chargino contributions 
to the $\varepsilon'/\varepsilon$ and also the total contribution
versus the imaginary part of the mass insertion $(\delta^d_{12})_{LL}$.  
As explained above, there are two sources for the gluino contributions
to $\varepsilon'/\varepsilon$: the usual $\Delta I=1/2$ chromomagnetic
dipole operator and the new $\Delta I=3/2$ penguin operators.  
Additionally, there are two sources for the chargino contribution to 
$\varepsilon'/\varepsilon$: the usual gluon and electroweak penguin
diagrams with single mass insertion and the contribution from the SUSY
effective $\bar{s}dZ$ vertex. As can be seen from this figure, the dominant
contribution to  $\varepsilon'/\varepsilon$ is due to the chargino 
exchange with one mass insertion. It turns out that the chargino 
contribution with two mass insertions is negligible. 
As expected the gluino contribution via the chromomagnetic 
operator is also negligible due to the severe cancellation between the LR and RL
contributions. The contribution from the $\Delta I=3/2$ operators
does not lead to significant results for $\varepsilon'/\varepsilon$. It
even becomes negative (opposite to the other contributions) for 
$\mathrm{Im}(\delta^d_{12})_{LL} > 2.5 \times 10^{-3}$. 

From this figure we conclude that a non--universality among the soft
scalar masses is necessary to get large values of 
$\varepsilon'/\varepsilon$ and $\varepsilon_K$. 

\section{CP violation in the $B$--system}
\subsection{Flavour--dependent SUSY phases and CP asymmetry in $B\to J/\psi 
K_s$}

Recent results from the $B$--factories have confirmed, for the first time,  
the existence of CP violation in the $B$--meason decays. In particular, 
the measurements of the CP asymmetry in the $B_d \to \psi K_s$ decay~
\cite{babar} have verified that CP is 
significantly violated in the 
$B$--sector. 
In $B_d - \bar{B}_d$ mixing, the flavor eigenstates are $B_d = (\bar{b} d)$ 
and $\bar{B}_d = (b \bar{d})$. 

It is 
customary to denote the corresponding 
mass eigenstates by $B_H = p B_d + q \bar{B}_d$
and $B_L = p B_d - q \bar{B}_d$ where indices H and L 
refer to heavy and light mass eigenstates
respectively, and 
$ p=(1+\bar{\varepsilon}_B)/\sqrt{2(1+|\bar{\varepsilon}_B|)}, ~~
q=(1-\bar{\varepsilon}_B)/\sqrt{2(1+|\bar{\varepsilon}_B|)}$
where $\bar{\varepsilon}_B$ is the corresponding 
CP violating parameter in the $B_d -\bar{B}_d$ system, analogous of
$\bar{\varepsilon}$ in the kaon system \cite{buras:2001}.
Then the strength of $B_d - \bar{B}_d$ mixing is described by
the mass difference 
\be
\Delta M_{B_d} = M_{B_H} - M_{B_L}.
\ee
whose present experimental value is 
$\Delta M_{B_d} = 0.484 \pm 0.010~\mathrm{(ps)}^{-1}$~\cite{buras:2001}.

The CP asymmetry of the  $B_d$ and $\bar{B}_d$ meson decay to the CP 
eigenstate $\psi K_S$ is given by
\be 
a_{\psi K_S}(t) = \frac{\Gamma(B^0_d(t) \to \psi K_S) -
\Gamma(\bar{B}^0_d(t) \to \psi K_S)}{\Gamma(B^0_d(t) \to \psi K_S) +
\Gamma(\bar{B}^0_d(t) \to \psi K_S)}=- a_{\psi K_S} \sin(\Delta
m_{B_d} t).
\ee
The most recent measurements of this asymmetry are given by~\cite{babar}
\bea
a_{\psi K_S}&=& 0.59 \pm 0.14 \pm 0.05 \quad \quad
\mathrm{(BaBar)} \;, \nonumber \\
a_{\psi K_S}&=& 0.99 \pm 0.14 \pm 0.06 \quad \quad
\mathrm{(Belle)} \;.
\label{2beta:bounds}
\eea
where the second and third numbers correspond to statistically and systematic 
errors respectively, and so the present world average is given by 
$a_{\psi K_S}=0.79\pm 12$.
These results show that there is a large CP asymmetry 
in the $B$ meson system. This implies that either the CP is not an approximate 
symmetry in nature and that the CKM mechanism is the dominant source of 
CP violation~\cite{Khalil:2002jq} or CP is an approximate symmetry with large 
flavour structure beyond the standard CKM matrix~\cite{branco}. 
Generally, $\Delta M_{B_d}$ and $a_{\psi K_S}$
can be calculated via
\bea
&&\Delta M_{B_d} = 2 
\vert \langle B^0_d \vert H_{\mathrm{eff}}^{\Delta B=2} \vert 
\bar{B}^0_d \rangle \vert \;, 
\label{DeltaMB}
\\
&&a_{\psi K_S} = \sin 2 \beta_{\mathrm{eff}} \;,~ \mathrm{and}~~~~ 
\beta_{\mathrm{eff}} = \frac{1}{2}\mathrm{arg} \langle B^0_d \vert 
H_{\mathrm{eff}}^{\Delta B=2} \vert \bar{B}^0_d \rangle \;.
\label{sin2b}
\eea
where $H_{\mathrm{eff}}^{\Delta B=2}$ is the effective Hamiltonian 
responsible of the ${\Delta B=2}$ transitions.
In the framework of the standard model (SM), 
$a_{\psi K_S}$ can be easily related 
to one of the inner angles of the unitarity triangles and 
parametrized by the $V_{CKM}$ elements as follows
\be
a_{\psi K_S}^{\mathrm{SM}} = \sin 2 \beta,~~ \beta = \mathrm{arg}\left(-\frac{V_{cd}
V_{cb}^*}{V_{td} V_{tb}^*} \right),
\ee

In supersymmetric theories the effective Hamiltonian 
for $\Delta B= 2$ transitions, can be generated,
in addition to the $W$ box diagrams of SM,
through other box diagrams mediated by charged Higgs, neutralino, 
gluino, and chargino exchanges. The 
Higgs contributions are suppressed by the quark masses and can be neglected. 
The neutralino exchange diagrams are also very suppressed compared 
to the gluino and chargino ones, due to their electroweak
neutral couplings to fermion and sfermions.
Thus, the dominant SUSY contributions to the 
off diagonal entry in the $B$-meson mass matrix, $\mathcal{M}_{12}(B_d) = 
\langle B^0_d \vert H_{\mathrm{eff}}^{\Delta B=2} \vert \bar{B}^0_d \rangle$,
is given by
\begin{equation}
\mathcal{M}_{12}(B_d) = \mathcal{M}_{12}^{\mathrm{SM}}(B_d) + 
\mathcal{M}_{12}^{\tilde{g}}(B_d) + \mathcal{M}_{12}^{\tilde{\chi}^+}(B_d).
\end{equation}
where $\mathcal{M}_{12}^{\mathrm{SM}}(B_d)$, 
$\mathcal{M}_{12}^{\tilde{g}}(B_d)$, and $\mathcal{M}_{12}^{\tilde{\chi}^+}$
indicate the SM, gluino, and chargino contributions respectively.
The SM contribution 
is known at NLO accuracy in QCD \cite{buras:2001} 
(as well as the leading SUSY contributions \cite{gluinoB})
and it is given by
\be
\mathcal{M}_{12}^{\mathrm{SM}}(B_d)= \frac{G_F^2}{12 \pi^2} \eta_{B} 
\hat{B}_{B_d} f^2_{B_d}
M_{B_d} M_W^2 (V_{td} V_{tb}^*)^2 S_0(x_t), 
\ee
where $f_{B_d}$ is the B meson decay constant, $\hat{B}_{B_d}$ is
the renormalization group invariant $B$ parameter (for its
definition and numerical value, see Ref. \cite{buras:2001}
and reference therein) and $\eta=0.55\pm 0.01$. The function $S_0(x_t)$,
connected to the $\Delta B=2$ box diagram with $W$ exchange, is given by
\be
S_0(x_t) = \frac{4 x_t - 11 x_t^2 +x_t^3}{4(1-x_t)^2} - 
\frac{3 x_t^3 \ln x_t}{2(1-x_t)^3},
\ee
where $x_t= M_t^2/M_W^2$. 

The effect of supersymmetry can be simply
described by a dimensionless parameter $r_d^2$ and a phase $2 \theta_d$
defined as follows
\begin{equation}
r_d^2 e^{2 i \theta_d} = \frac{\mathcal{M}_{12}(B_d)}{\mathcal{M}_{12}^{\mathrm{SM}}
(B_d)},
\end{equation}
where 
$\Delta M_{B_d} = 2 \vert \mathcal{M}_{12}^{\mathrm{SM}}(B_d)\vert r_d^2$.
Thus, in the presence of SUSY contributions, the
CP asymmetry $B_d \to \psi K_S$ is modified, and now we have 
\be
a_{\psi K_S} = \sin 2 \beta_{\mathrm{eff}} = \sin (2 \beta + 2 \theta_d)\;.
\ee
Therefore, the measurement of $a_{\psi K_S}$ would not determine $\sin 2 \beta$ but 
rather $\sin 2 \beta_{\mathrm{eff}}$, where
\begin{equation}
2 \theta_d = \mathrm{arg}\left(1+\frac{\mathcal{M}_{12}^{\mathrm{SUSY}}(B_d)}
{\mathcal{M}_{12}^{\mathrm{SM}}(B_d)}\right),
\end{equation}
and $\mathcal{M}_{12}^{\mathrm{SUSY}}(B_d)= \mathcal{M}_{12}^{\tilde{g}}(B_d) + 
\mathcal{M}_{12}^{\tilde{\chi}^+}(B_d)$.

The most general effective Hamiltonian for $\Delta B=2$ 
processes, induced by gluino and chargino exchanges through 
$\Delta B=2$ box diagrams, can be expressed as
\be
H^{\Delta B=2}_{\rm eff}=\sum_{i=1}^5 C_i(\mu) Q_i(\mu) + 
\sum_{i=1}^3 \tilde{C}_i(\mu) \tilde{Q}_i(\mu) + h.c. \;,
\label{Heff}
\ee
where $C_i(\mu)$, $\tilde{C}_i(\mu)$ and $Q_i(\mu)$, $\tilde{Q}_i(\mu)$
are the Wilson coefficients and operators respectively 
renormalized at the scale $\mu$, with
\bea 
Q_1 &=&\bar{d}_L^{\alpha} \gamma_{\mu} b_L^{\alpha}~ 
\bar{d}_L^{\beta} \gamma_{\mu} b_L^{\beta},~~~~
Q_2 = \bar{d}_R^{\alpha} b_L^{\alpha}~ 
\bar{d}_R^{\beta} b_L^{\beta},~~~~ Q_3 = \bar{d}_R^{\alpha} b_L^{\beta}~ 
\bar{d}_R^{\beta} b_L^{\alpha},\nonumber\\
Q_4 &=& \bar{d}_R^{\alpha} b_L^{\alpha}~ \bar{d}_L^{\beta} 
b_R^{\beta},~~~~~~~~~
Q_5 = \bar{d}_R^{\alpha} b_L^{\beta}~ \bar{d}_L^{\beta} b_R^{\alpha}\, .
\label{operators}
\eea
In addition, the operators $\tilde{Q}_{1,2,3}$ are obtained from $Q_{1,2,3}$ 
by exchanging $L \leftrightarrow R$. 
In the case of the gluino exchange all 
the above operators give significant contributions 
and the corresponding Wilson coefficients are given by \cite{gluinoB} 
\bea
C_1(M_S) &=& -\frac{\alpha_s^2}{216 m_{\tilde{q}}^2} \left(24x f_6(x)+
66\tilde{f}_6(x)
\right) (\delta_{13}^d)_{LL},\nonumber\\
C_2(M_S) &=& -\frac{\alpha_s^2}{216 m_{\tilde{q}}^2} 204 x  f_6(x) 
(\delta_{13}^d)_{RL},\nonumber\\
C_3(M_S) &=& \frac{\alpha_s^2}{216 m_{\tilde{q}}^2} 36 x  f_6(x) 
(\delta_{13}^d)_{RL},\\
C_4(M_S) &=& -\frac{\alpha_s^2}{216 m_{\tilde{q}}^2} \left[\left(504 x f_6(x) - 72 
\tilde{f}_6(x)\right) (\delta_{13}^d)_{LL}(\delta_{13}^d)_{RR}
-132 \tilde{f}_6(x) (\delta_{13}^d)_{LR}(\delta_{13}^d)_{RL}\right] ,
\label{wilsong}\nonumber\\
C_5(M_S) &=& -\frac{\alpha_s^2}{216 m_{\tilde{q}}^2} \left[\left(24 x f_6(x) +120 
\tilde{f}_6(x)\right) (\delta_{13}^d)_{LL}(\delta_{13}^d)_{RR}
-180 \tilde{f}_6(x) (\delta_{13}^d)_{LR}(\delta_{13}^d)_{RL}\right],\nonumber
\eea
where $x=m^2_{\tilde{g}}/m^2_{\tilde{q}}$ and $\tilde{m}^2$ is 
an average squark mass.
The expression for the 
functions $f_6(x)$ and $\tilde{f}_6(x)$ can be found in Ref.\cite{gluinoB}. 
The Wilson coefficients 
$\tilde{C}_{1,2,3}$ are simply obtained by 
interchanging $L\leftrightarrow R$ in the mass insertions appearing in 
$C_{1,2,3}$. 

Now we considerthe chargino contribution to the 
effective Hamiltonian in Eq.(\ref{Heff}) in the mass insertion approximation.
The dominant chargino exchange, can give significant
contributions only to the operators 
$Q_1$ and $Q_3$ in Eq.(\ref{operators})~\cite{emidio3}.
At the first order in the mass insertion approximation,
the Wilson coefficients $C_{1,3}^{\chi}(M_S)$ (by taking different
the mass of the stop--right from the average squark mass) take the 
form~\cite{emidio3}
\bea
C_1^{\chi}(M_S) &=& \frac{g^4}{768 \pi^2 \tilde{m}^2} 
\sum_{i,j}
\Big\{\vert V_{i1}\vert^2 \vert V_{j1} \vert^2\left( 
(\delta^u_{LL})_{31}^2 + 2\lambda (\delta^u_{LL})_{31}
(\delta^u_{LL})_{32}\right) L_2(x_i,x_j)
\nonumber \\ 
&-& 2 Y_t  
\vert V_{i1}\vert^2 V_{j1} V_{j2}^*\Big(
(\delta^u_{LL})_{31} (\delta^u_{RL})_{31}+\lambda
(\delta^u_{LL})_{32} (\delta^u_{RL})_{31}+\lambda
(\delta^u_{LL})_{31} (\delta^u_{RL})_{32}\Big) R_2(x_i,x_j,z)
\nonumber \\
&+& Y_t^2 V_{i1} V_{i2}^{*} V_{j1} V_{j2}^{*}\Big(
(\delta^u_{RL})_{31}^2 +2\lambda
(\delta^u_{RL})_{31} (\delta^u_{RL})_{32}~ \Big) \tilde{R}_2(x_i,x_j,z)
\, \Big\},
\label{C1stop}
\\
C_3^{\chi}(M_S) &=& \frac{g^4\, Y_b^2}{192 \pi^2 \tilde{m}^2}
\sum_{i,j}
U_{i2} U_{j2} V_{j1} V_{i1}~ \Big( (\delta^u_{LL})_{31}^2 + 
2 \lambda  (\delta^u_{LL})_{31} (\delta^u_{LL})_{32} \Big) 
L_0(x_i, x_j),
\label{C3}
\eea
where 
$x_i= m^2_{\tilde{\chi}_i^+}/\tilde{m}^2$, and the functions $L_0(x,y)$
and $L_2(x,y)$ are given by 
\bea
L_0(x,y)&=&\sqrt{xy}\left(\frac{x\, h_0(x)- y\, h_0(y)}{x-y}\right),
~~
h_0(x)=\Frac{-11+7x-2x^2}{(1-x)^3}- \Frac{6 \ln x}{(1-x)^4}
\nonumber \\
L_2(x,y) &=& \Frac{x\, h_2(x) -y\, h_2(y)}{x-y},~~~~~~~~~~~~~
h_2(x)=\Frac{2+5 x -x^2}{(1-x)^3}
+ \Frac{6 x \ln x}{(1-x)^4}
\eea
and
\bea
R_2(x,y,z)&=&\frac{1}{x-y}\left(H_2(x,z)-H_2(y,z)\right),~~~
\tilde{R}_2(x,y,z)=\frac{1}{x-y}\left(\tilde{H}_2(x,z)-
\tilde{H}_2(y,z)\right)
\nonumber \\
H_2(x,z)&=&\frac{3}{D_2(x,z)} \Big\{ (-1 + x) ( x - z)( -1 + z) 
( -1 - x - z + 3 x z )\nonumber\\
&+& 6 x^2 ( -1 + z)^3 \log(x) - 6 ( -1 + x)^3 z^2 \log (z) \Big\}
\nonumber \\
\tilde{H}_2(x,z)&=&\frac{-6}{\tilde{D}_2(x,z)} \Big\{ (-1+x) ( x - z) (-1 + z)
( x + ( -2 + x ) z) \nonumber\\
&+& 6 x^2 ( -1 + z )^3 \log (x) - 6 (-1+x)^2 z (-2 x + z + z^2) \log (z) \Big\}
\eea
Here
$D_2(x,z)={{{\left( -1+x \right) }^3}\,
     \left( x - z \right) \,{{\left(-1 + z \right) }^3}}$
and
$\tilde{D}_2(x,z)={{\left( -1 + x\right) }^2}\,
     {{\left( x - z \right) }^2}\,{{\left( -1 + z \right) }^3}$.
Notice that in the limit $z\to 1$, both the functions 
$R_2(x,y,z)$ and $\tilde{R}_2(x,y,z)$ tend to $L_2(x,y)$.
As in the gluino case, 
the corresponding results for $\tilde{C_1}$ and $\tilde{C_1}$ coefficients 
are simply obtained by interchanging $L\leftrightarrow R$ in 
the mass insertions appearing in the expressions for $C_{1,3}$.

In order to connect $C_i(M_S)$ at SUSY scale $M_S$
with the corresponding low energy ones
$C_i(\mu)$ (where $ \mu\simeq {\cal O}(m_b)$),
one has to solve the renormalization group 
equations (RGE) for the Wilson coefficients 
corresponding to the effective Hamiltonian in (\ref{Heff}).
Then, $C_i(\mu)$ will be related to $C_i(M_S)$
by the following relation\cite{gluinoB}
\be
C_r(\mu) = \sum_i \sum_s \left( b_i^{(r,s)} + 
\eta c_{i}^{(r,s)}\right) \eta^{a_i} C_s(M_S),
\label{CWnlo}
\ee
where $M_S>m_t$ and $\eta=\alpha_S(M_S)/\alpha_S(\mu)$.
The values of the coefficients $b_i^{(r,s)}$, $c_i^{(r,s)}$, and 
$a_i$ appearing in (\ref{CWnlo}) can be found in Ref.\cite{gluinoB}.

In tables (\ref{table10}) and 
(\ref{table11}) we present the bounds on real and imaginary parts of mass 
insertions respectively, by taking into account a light stop--right 
mass\cite{emidio3}. We considered two
representative cases of $\tilde{m}_{t_R}=100, ~200$ GeV.
Clearly, the light stop--right effect does not affect bounds on mass 
insertions containing LL interactions. 
From these results we could see that the effect of taking 
$\tilde{m}_{t_R} < \tilde{m}$ is sizable. In particular, on the bounds 
of the mass insertions $(\delta_{RL}^u)_{31}(\delta_{RL}^u)_{3i}$,
$(i=1,2)$ which are the most sensitive to a light stop--right.

From the results in tables (10) and (11), it is remarkable 
to notice that,
in the limit of very heavy squark masses but with fixed right stop and 
chargino masses, the bounds on the mass insertions $(\delta^u_{RL})_{31} 
(\delta^u_{RL})_{3i}$ tend to constant values. This is indeed an interesting 
property which shows a particular non--decoupling 
effect of supersymmetry when two 
light right--stop run inside the diagrams in Fig. (1).  This feature is 
related to the infrared singularity of the loop function 
$\tilde{R}_2(x,x,z)$ in the limit $z \to 0$. 
In particular, we find that $\lim_{z\to 0} \tilde{R}_2(x,x,z)=f(x)/x$, where
$x=m_{\chi}^2/\tilde{m}^2$, and $f(x)$ is a non-singular and non-null function
in $x=0$. Then, in the limit $\tilde{m} >> m_{\chi}$ the
 rescaling factor $1/\tilde{m}^2$ in $C_1^{\chi}$ will be canceled by the 
$1/x$ dependence in the loop function and replaced by $1/m_{\chi}^2$
times a constant factor.

This is a quite interesting result, since it shows that in the
case of light right stop and charginos masses, in comparison to the other 
squark masses, the SUSY contribution (mediated by charginos) to the 
$\Delta B=2$ processes might not decouple and could be sizable, 
provided that the mass insertions $(\delta^u_{RL})_{3i}$ are large enough.
This effect could be achieved, for instance, in supersymmetric models with
non--universal soft breaking terms. 

\begin{table}[h]
\begin{center}
\begin{tabular}{|c|c||c|c|c|}
\hline
$\tilde{m}$
& 
$\tilde{m}_{t_R}$
& 
$\sqrt{\bigl\vert{\rm Re}\left[(\delta_{RL}^u)^2_{31}\right]\bigr\vert}$ 
&  
$\sqrt{\bigl\vert{\rm Re}
\left[(\delta_{LL}^u)_{31}(\delta_{RL}^u)_{3i}\right]\bigr\vert}$ 
& 
$\sqrt{\bigl\vert{\rm Re}\left[(\delta_{RL}^u)_{31}(\delta_{RL}^u)_{32}
\right]\bigr\vert}$ 
\\
\hline
\hline
400 & 100 & 1.9$\times 10^{-1}$ & 1.6(3.3)$\times 10^{-1}$ & 2.8$\times 10^{-1}$ \\
\hline
600 & 100 & 1.8$\times 10^{-1}$ & 1.9(4.0)$\times 10^{-1}$ & 2.6$\times 10^{-1}$ \\
\hline
800 & 100 & 1.8$\times 10^{-1}$ & 2.3(4.9)$\times 10^{-1}$ & 2.6$\times 10^{-1}$ \\
\hline
\hline
400 & 200 & 3.5$\times 10^{-1}$ & 2.0(4.2)$\times 10^{-1}$ & 5.2$\times 10^{-1}$ \\
\hline
600 & 200 & 3.3$\times 10^{-1}$ & 2.3(5.0)$\times 10^{-1}$ & 4.9$\times 10^{-1}$ \\
\hline
800 & 200 & 3.2$\times 10^{-1}$ & 2.8(5.9)$\times 10^{-1}$ & 4.8$\times 10^{-1}$\\
\hline
\end{tabular}
\end{center}
\caption{Upper bounds on real parts of mass insertions 
as in tables (\ref{table3})--(\ref{table4}), for some values 
of $\tilde{m}$ and $\tilde{m}_{t_R}$ (in GeV). 
In the fourth column the first number and the one in parenthesis correspond 
to $i=1$ and $i=2$ respectively.
Upper bounds on mass insertions involving only LL interactions
are the same as in tables (\ref{table3})--(\ref{table4}).}
\label{table10}
\end{table}


\begin{table}[h]
\begin{center}
\begin{tabular}{|c|c||c|c|c|}
\hline
$\tilde{m}$
& 
$\tilde{m}_{t_R}$
& 
$\sqrt{\bigl\vert{\rm Im}\left[(\delta_{RL}^u)^2_{31}\right]\bigr\vert}$ 
&  
$\sqrt{\bigl\vert{\rm Im}
\left[(\delta_{LL}^u)_{31}(\delta_{RL}^u)_{3i}\right]\bigr\vert}$ 
& 
$\sqrt{\bigl\vert{\rm Im}\left[(\delta_{RL}^u)_{31}(\delta_{RL}^u)_{32}
\right]\bigr\vert}$ 
\\
\hline
\hline
400 & 100 & 2.1$\times 10^{-1}$ & 1.8(3.7)$\times 10^{-1}$ & 3.1$\times 10^{-1}$ \\
\hline
600 & 100 & 2.0$\times 10^{-1}$ & 2.2(4.6)$\times 10^{-1}$ & 3.0$\times 10^{-1}$ \\
\hline
800 & 100 & 2.0$\times 10^{-1}$ & 2.6(5.5)$\times 10^{-1}$ & 3.0$\times 10^{-1}$ \\
\hline
\hline
400 & 200 & 4.0$\times 10^{-1}$ & 2.2(4.8)$\times 10^{-1}$ & 6.0$\times 10^{-1}$ \\
\hline
600 & 200 & 3.7$\times 10^{-1}$ & 2.7(5.6)$\times 10^{-1}$ & 5.6$\times 10^{-1}$ \\
\hline
800 & 200 & 3.6$\times 10^{-1}$ & 3.1(6.7)$\times 10^{-1}$ & 5.4$\times 10^{-1}$\\
\hline
\end{tabular}
\end{center}
\caption{
Upper bounds on imaginary parts of mass insertions 
as in tables (\ref{table5})--(\ref{table6}), for some values
of $\tilde{m}$ and $\tilde{m}_{t_R}$ $\tilde{m}$ (in GeV).
In the fourth column the first number and the one in parenthesis correspond 
to $i=1$ and $i=2$ respectively.
Upper bounds on mass insertions involving only LL interactions
are the same as in tables (\ref{table5})--(\ref{table6}).}
\label{table11}
\end{table}

In the case of SUSY models with Hermitian flavor structure,
we find that in most of the parameter space the chargino gives the 
dominant contribution to $B_d - \bar{B}_d$ mixing and the CP 
asymmetry $a_{J/\psi K_S}$ and the gluino is sub-dominant. 
As we emphasized above, 
in order to have a significant gluino contribution for $\tilde{m}\sim m_g\sim
500$ GeV ({\it i.e.}, $m_0\sim M_{1/2}\sim 200$ at GUT scale), the real and 
imaginary parts of mass insertion 
$(\delta^d_{13})_{LL}$ or $(\delta^d_{13})_{LR}$ 
should be of order $10^{-1}$ and $10^{-2}$ respectively. 
However, with the above hierarchical Yukawas we find 
that these mass insertions are two orders of magnitude below the required 
values so that the gluino contributions are very small.

Concerning the chargino amplitude to the CP asymmetry $a_{J/\psi K_S}$,
we find that the mass insertions 
$(\delta^u_{31})_{RL}$ and $(\delta^u_{31})_{LL}$
give the leading contribution to $a_{J/\psi K_S}$.
 However, for 
the representative case of $m_0 = m_{1/2} =200$ and $\phi_{ij}
\simeq \pi/2$ the values of these mass insertions are given by
\begin{eqnarray}
\sqrt{\vert \mathrm{Im}[(\delta^u_{LL})_{31}]^2\vert}&=&6\times 10^{-4},\\
\sqrt{\vert \mathrm{Im}[(\delta^u_{LL})_{31}]^2\vert}&=&4\times 10^{-3},\\
\sqrt{\vert \mathrm{Im}[(\delta^u_{LL})_{31} (\delta^u_{RL})_{32}]\vert}&=&
1\times 10^{-4}.
\end{eqnarray}
These results show that, also for this class of models,
SUSY contributions cannot give sizable 
effects to $a_{J/\psi K_S}$.  
As expected, with hierarchical Yukawa couplings 
(where the mixing between different generations is very small), 
the SUSY contributions to the $B-\bar{B}$ mixing and the CP asymmetry of 
$B_d \to J/\psi K_S$ are sub-dominant and the SM should give the dominant 
contribution.
%
\subsection{Flavour--dependent SUSY phases and CP asymmetry in $B \to
X_s \gamma$ decays}

The Standard Model prediction for the CP asymmetry in rare $B$ decays
$B \to X_s \gamma$ is very small, less than
$1\%$. Thus, the observation of sizeable asymmetry in this decay  
would be a clean signal of new physics. The most recent result 
reported by CLEO collaboration for the CP asummetry in  these decays is 
\cite{cleo}
\begin{equation}
-9\% < A_{CP}^{b\to s \gamma}  < 42 \%~,
\end{equation}
and it is expected that the measurements of $A_{CP}^{b\to s \gamma}$ 
will be improved in the next few years at the $B$--factories.

Supersymmetric predictions for $A_{CP}^{b\to s \gamma}$ are strongly
dependent on the flavour structure of the soft breaking terms. 
It was shown that in the universal case, as in the minimal supergravity 
models, the prediction of the asymmetry is less than $2\%$, since in this 
case the EDM of the electron and neutron constrain
the SUSY CP--violating phases to be very small~\cite{susyB}. 
Furthermore, it is also known that in this case, one can not get any 
sizeable SUSY contribution to the CP--violating observables, $\varepsilon$ and
$\varepsilon'/\varepsilon$.

In this subsection, we explore the effect of these large flavour--dependent 
phases on inducing a direct CP violation in $B \to X_s \gamma$ decay. We 
will show that the values of the asymmetry $A_{CP}^{b\to s \gamma}$ in this 
class of models are much larger than the SM prediction in a wide region of 
the parameter space allowed by experiments, namely the EDM experimental 
limits and the bounds on the branching ratio of $B \to X_s \gamma$. The 
enhancement of $A_{CP}^{b\to s \gamma}$ is due to the
important contributions from gluino--mediated diagrams, in this
scenario, in addition to the usual chargino and charged Higgs contributions.

The relevant operators for $b \to s \gamma$ decay are given by
\bea
Q_2 &=& \bar{s}_L \gamma^{\mu} c_L~ \bar{c}_L \gamma_{\mu} b_L, \no
Q_7 &=& \Frac{e}{16 \pi^2} m_b \bar{s}_L \sigma^{\mu \nu} b_R F_{\mu \nu}, \no
Q_8 &=& \Frac{g_s}{16 \pi^2} m_b \bar{s}_L \sigma^{\mu \nu} T^a b_R G^a_{\mu \nu}.
\eea

The expression for the asymmetry  $A_{CP}^{b\to s \gamma}$, corrected to 
next--to--leading order is given by \cite{susyB}
\bea
A_{CP}^{b\to s \gamma} &=& \Frac{4\alpha_s(m_b)}{9 \vert C_7 \vert^2} 
\biggl\{\biggl[\frac{10}{9} - 2 z~ (v(z)+b(z,\delta))\biggr] Im[C_2 C_7^*] 
+ Im[C_7 C^*_8]\no &+& \frac{2}{3} z~b(z,\delta) Im[C_2 C_8^8] \biggr \}, 
\label{asymmetry}
\eea
where $z=m_c^2/m_b^2$. The functions $v(z)$ and $b(z,\delta)$ can be found in
Ref.\cite{susyB}. The parameter $\delta$ is related to the experimental 
cut on the photon
energy, $E_{\gamma} > (1-\delta) m_b/2$, which is assumed to be 0.9.

The SUSY contributions to the Wilson coefficients
$C_{7,8}$ are obtained by calculating the $b\to s \gamma$ and $b \to s g $
amplitudes at the electroweak scale respectively. 
The leading--order contributions to these amplitudes are given by the 1--loop
magnetic-dipole and chromomagnetic dipole penguin diagrams respectively,
mediated by charged Higgs boson, chargino, gluino, and neutralino exchanges. 
As pointed out above, SUSY models with non--universal $A$--
terms may induce non--negligible contributions
to the dipole operators $\tilde{Q}_{7,8}$ which have opposite chirality to
$Q_{7,8}$. In the MSSM these contributions are suppressed by terms of order 
${\cal O}(m_s/m_b)$ due to the universality of the $A$--terms. However, in 
our case we should take them into account. Denoting by $\tilde{C}_{7,8}$ the 
Wilson coefficients multiplying the new operators $\tilde{Q}_{7,8}$ the 
expression for the asymmetry in Eq.(\ref{asymmetry}) will be modified by 
making the replacement 
\be
C_i C_j^* \to C_i C_j^* + \tilde{C}_i \tilde{C}_j^*.
\label{chirality}
\ee

The expressions for $\tilde{C}_{7,8}$ are given in the appendix 
and $\tilde{C}_2=0$ (there is no operator similar to $Q_2$
containing right--handed quark fields).
Note that including these modifications (\ref{chirality})
may enhance the branching ratio of $B \to X_s \gamma$ and reduce the CP 
asymmetry, since $\vert C_7 \vert^2$ is replaced by $\vert C_7 \vert^2 + 
\vert \tilde{C}_7 \vert^2$ in the denominator of Eq.(\ref{asymmetry}). 
If so, neglecting this contribution could lead to an incorrect conclusion. 

\begin{figure}[h]
\centerline{  
\epsfig{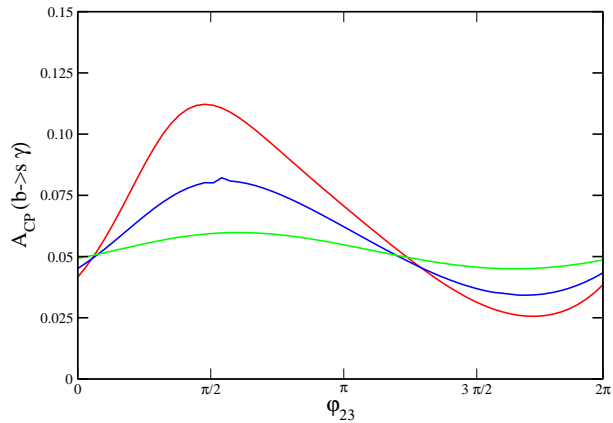}
} 
\caption{CP asymmetry $A_{CP}^{b\to s \gamma}$ as a
function of the  flavor-dependent phase $\phi_{23}$  
for $m_0 \simeq 150$ GeV, $\tan \beta =10$, and $m_{\chi^+}\simeq 100$ GeV,
$\vert A_{ii} \vert =m_0$. Curve 1: $\vert A_{ij}\vert=m_0 (i \neq j)$; 
curve 2: $\vert A_{ij}\vert=3m_0$;curve 3: $\vert A_{ij}\vert=5m_0$. 
}
\label{cpasy}
\end{figure}

In the EDM-free models we are considering, we found that 
the flavor dependent phase $\phi_{23}$ gives a large contribution 
to the CP asymmetry. This casn simply be understood by using the mass
insertion; the gluino contributions to $C_7$ and $C_8$ (and also
$\tilde{C}_7$ and $\tilde{C}_8$) are proportional to $(\delta_{23}^d)_{LR}$
which receives a dominant contributions from $A_{23}$ entry. The effect 
of the other flavor dependent phases on $A_{CP}^{b \to s \gamma}$ is 
found to be very small. 
In Fig. \ref{cpasy} we show the dependence of 
$A_{CP}^{b \to s \gamma}$ on the phase $\phi_{23}$ in the case of hermitian 
$A$-terms. We assume $\tan \beta = 10$, the diagonal elements $A_{ii}=m_0$,
and consider three values for the off-diagonal elements $A_{ij}, (i\neq j)$: 
$\vert A_{ij} \vert = m_0$ (curve 1), 
$\vert A_{ij} \vert = 3 m_0$ (curve 2), and 
$\vert A_{ij} \vert = 5 m_0$ (curve 3).  The conditions of the correct 
branching ratio has been automatically imposed. 

We see that the CP asymmetry $A_{CP}^{b\to s \gamma}$ can be as large
as $15\%$, which can be accessible at the $B$ facrories. Also  as 
emphasised in Ref.\cite{bailin}, larger $\tan \beta$ is, the larger the 
CP asymmetry $A_{CP}^{b\to s \gamma}$ become. In the case of symmetric 
$A$-terms, the CP phase in $(\delta^d_{23})_{LR}$ is due to the CKM
mixing only, however, it implies a considerable 
CP asymmetry, $A_{CP}^{b\to s \gamma} \sim \mathcal{O} (10\%)$



\begin{thebibliography}{99}

\bibitem{edmexp}
 P.G. Harris {\it et al.}, Phys. Rev. Lett. {\bf 82} (1999), 904;
 see also the discussion in S.K. Lamoreaux and R. Golub, Phys. Rev.  {\bf D61} (2000),
 051301; E.D. Commins {\it et al.}, Phys. Rev. {\bf A50} (1994), 2960;
 M.V. Romalis, W.C. Griffith, and E.N. Fortson,
 Phys.\ Rev.\ Lett.\ {\bf 86}, 2505 (2001);
 J.P. Jacobs {\it et al.}, Phys. Rev. Lett. {\bf 71} (1993), 3782.
\bibitem{AKL}
 S.~Abel, S.~Khalil and O.~Lebedev,
 Nucl.\ Phys.\ B {\bf 606}, 151 (2001), and reference therein.
\bibitem{susyK}
S.~Abel and J.~Frere, \prd{55}{1997}{1623};
A. Masiero and H. Murayama, \prl{83}{1999}{907};
S.~Khalil, T.~Kobayashi, and A.~Masiero, \prd{60}{1999}{075003};
S.~Khalil and T.~Kobayashi, \plb{460}{1999}{341};
S.~Khalil, T.~Kobayashi and O.~Vives, Nucl.\ Phys.\ B {\bf 580}, 275 (2000);
M. Brhlik, L. Everett, G. L. Kane, S. F. King, and O. Lebedev,
\prl{84}{2000}{3041};
R.~Barbieri, R.~Contino, and A.~Strumia, arXiv:hep-ph/9908255;
K.~Babu, B.~Dutta, and R.N.~Mohapatra, \prd{61}{2000}{091701}.
\bibitem{susyB}
M.~Aoki, G.~C.~Cho and N.~Oshimo,
Nucl.\ Phys.\ B {\bf 554}, 50 (1999);
A.~L.~Kagan and M.~Neubert,
Eur.\ Phys.\ J.\ C {\bf 7}, 5 (1999);
Phys.\ Rev.\ D {\bf 58}, 094012 (1998);
D.~Bailin and S.~Khalil, Phys.\ Rev.\ Lett.\  {\bf 86}, 4227 (2001).
\bibitem{hooft}
G. 't Hooft, in {\it Recent Advances in Gauge Theories},
Proceedings of the Cargese Summer Institute, Cargese, France, 1979, edited by
G. t' Hooft {\it et al.}, NATO Advanced Study Institute Series B; 
Physics Vol.59 (Plenum, New York, 1980).
\bibitem{sgravity}
H.~P.~Nilles, Phys.\ Rept.\  {\bf 110}, 1 (1984).
\bibitem{gmsb}
G.~F.~Giudice and R.~Rattazzi, Phys.\ Rept.\  {\bf 322}, 419 (1999), and reference therein.
\bibitem{dvali}
G.~R.~Dvali, G.~F.~Giudice and A.~Pomarol,
Nucl.\ Phys.\ B {\bf 478}, 31 (1996).
\bibitem{pospelov} T. Falk, K.A. Olive, M. Pospelov, R. Roiban, Nucl.\ Phys.\ {\bf B560} (1999), 3.
\bibitem{pospelov1} M. Pospelov and A. Ritz, Phys.\ Rev.\ D {\bf 63}, 073015 (2001).
\bibitem{khrip1} V.M. Khatsimovsky, I.B. Khriplovich and A.R. Zhitnitsky, Z. Phys. {\bf C36} (1987), 455;
         V.M. Khatsimovsky, I.B. Khriplovich and A.S. Yelkhovsky, Annals Phys. {\bf 186} (1988), 1;
         V.M. Khatsimovsky and  I.B. Khriplovich, Phys. Lett. {\bf B296} (1992), 219.
\bibitem{khrip} I.B. Khriplovich and S.K. Lamoreaux, {\it ``CP Violation Without
        Strangeness''}, Springer, 1997.
\bibitem{nda} H. Georgi and A. Manohar, Nucl. Phys. {\bf B234}, 189 (1984);
        R. Arnowitt, J.L. Lopez and D.V. Nanopoulos, Phys. Rev. {\bf D42} (1990),
         2423; R. Arnowitt, M.J. Duff and K.S. Stelle, Phys. Rev. {\bf D43} (1991), 3085.
\bibitem{nath} T. Ibrahim and P. Nath, Phys. Rev. {\bf D57} (1998), 478;
        Errata-{\it ibid.} {\bf D58}, 019901 (1998); {\it ibid.} {\bf D60}, 019901 (1999);
        Phys. Lett. {\bf B418}, 98 (1998); Phys. Rev.  {\bf D58}, 111301 (1998); Erratum-{\it ibid.} {\bf D60},
        099902 (1999).
\bibitem{ellis} J. Ellis, R. Flores, Phys. Lett. {\bf B377} (1996), 83.
\bibitem{bartl} A. Bartl, T. Gajdosik, W. Porod, P. Stockinger, and H. Stremnitzer,
        Phys. Rev. {\bf D60} (1999), 073003.
\bibitem{Weinberg:1989dx}
        S.~Weinberg, Phys.\ Rev.\ Lett.\ {\bf 63}, 2333 (1989);
        E.~Braaten, C.~Li and T.~Yuan, Phys.\ Rev.\ Lett.\ {\bf 64}, 1709 (1990).
\bibitem{Dai:1990xh}
       J.~Dai, H.~Dykstra, R.~G.~Leigh, S.~Paban and D.~Dicus,
        Phys.\ Lett.\ {\bf B237}, 216 (1990).
\bibitem{chang}
   D.~Chang, W.~Keung and A.~Pilaftsis,
   Phys.\ Rev.\ Lett.\ {\bf 82}, 900 (1999); Erratum-{\it ibid.} {\bf 83}, 3972 (1999).
\bibitem{nir} G.~Eyal and Y.~Nir, Nucl.\ Phys.\ {\bf B528} (1998), 21; and references
   therein.
\bibitem{babar}
B.~Aubert {\it et al.}  [BABAR Collaboration],
Phys.\ Rev.\ Lett.\  {\bf 87}, 091801 (2001);
K.~Abe {\it et al.}  [Belle Collaboration],
Phys.\ Rev.\ Lett.\  {\bf 87}, 091802 (2001);
T.~Affolder {\it et al.}  [CDF Collaboration],
Phys.\ Rev.\ D {\bf 61}, 072005 (2000).
\bibitem{heavy} P. Nath, Phys. Rev. Lett. {\bf 66} (1991), 2565; Y. Kizukuri and N. Oshimo,
        Phys. Rev. {\bf D46} (1992) 3025.
\bibitem{cancel} T. Falk and K.A. Olive, Phys. Lett. B {\bf 375}, 196 (1996);
         Phys. Lett. B {\bf 439}, 71 (1998);
         M. Brhlik, G.J. Good and G.L. Kane, Phys. Rev. D {\bf 59}, 115004 (1999).
\bibitem{savoy}  S. Pokorski, J. Rosiek and C.A. Savoy, Nucl. Phys. B {\bf 570}, 81 (2000).
\bibitem{Abel:2000hn} S.~Abel, D.~Bailin, S.~Khalil and O.~Lebedev, 
Phys.\ Lett.\ B {\bf 504}, 241 (2001).
\bibitem{Bertolini:1991if}
S.~Bertolini, F.~Borzumati, A.~Masiero and G.~Ridolfi,
Nucl.\ Phys.\ B {\bf 353}, 591 (1991).
\bibitem{stringcp}
S.~Abel, S.~Khalil and O.~Lebedev,
Phys.\ Rev.\ Lett.\  {\bf 89}, 121601 (2002).
\bibitem{Khalil:2002jq}
S.~Khalil,
arXiv:hep-ph/0202204.
\bibitem{CCB}
J.~M.~Frere, D.~R.~Jones and S.~Raby, Nucl.\ Phys.\ B {\bf 222}, 11 (1983);
L.~Alvarez-Gaume, J.~Polchinski and M.~B.~Wise,
Nucl.\ Phys.\ B {\bf 221}, 495 (1983);
J.~P.~Derendinger and C.~A.~Savoy,
Nucl.\ Phys.\ B {\bf 237}, 307 (1984);
C.~Kounnas, A.~B.~Lahanas, D.~V.~Nanopoulos and M.~Quiros,
Nucl.\ Phys.\ B {\bf 236}, 438 (1984);
J.~A.~Casas, A.~Lleyda and C.~Munoz,
Nucl.\ Phys.\ B {\bf 471}, 3 (1996).
\bibitem{casas}
J.~A.~Casas and S.~Dimopoulos,
Phys.\ Lett.\ B {\bf 387}, 107 (1996)
\bibitem{EH}
M.~Ciuchini {\it et al},
JHEP {\bf 10},008 (1998).
\bibitem{olegchar}
S.~Khalil and O.~Lebedev, Phys.\ Lett.\ B {\bf 515}, 387 (2001).
\bibitem{buras:2001}
A.~J.~Buras, hep-pph/0101336.
\bibitem{branco:book}
See, for example,
G.~C.~Branco, L.~Lavoura and J.~P.~Silva,
\bibitem{Gabbiani}
F.~Gabbiani, E.~Gabrielli, A.~Masiero and L.~Silvestrini,
Nucl.\ Phys.\ B {\bf 477}, 321 (1996).
\bibitem{ciuchini}
M.~Ciuchini {\it et al.}, JHEP {\bf 9810}, 008 (1998).
\bibitem{buras2}
A.~J.~Buras, M.~Jamin and M.~E.~Lautenbacher,
Nucl.\ Phys.\ B {\bf 408}, 209 (1993).
\bibitem{emidio}
S.~Bertolini, M.~Fabbrichesi and E.~Gabrielli,
Phys.\ Lett.\ B {\bf 327}, 136 (1994).
\bibitem{buras3}
A.~J.~Buras, P.~Gambino, M.~Gorbahn, S.~Jager and L.~Silvestrini,
Nucl.\ Phys.\ B {\bf 592}, 55 (2001).
\bibitem{neubert}
A.~L.~Kagan and M.~Neubert,
Phys.\ Rev.\ Lett.\  {\bf 83}, 4929 (1999).
\bibitem{branco}
G.~C.~Branco, M.~E.~Gomez, S.~Khalil and A.~M.~Teixeira,
arXiv:hep-ph/0204136.
\bibitem{gluinoB}
D.~Becirevic {\it et al.}, arXiv:hep-ph/0112303.
\bibitem{emidio3}
E.~Gabrielli and S.~Khalil, arXiv:hep-ph/0207288.
\bibitem{cleo}
T.~E.~Coan {\it et al.}  [CLEO Collaboration],
Phys.\ Rev.\ Lett.\  {\bf 86}, 5661 (2001)

















\end{thebibliography}
\end{document}